\documentclass[twoside,twocolumn,9pt]{article}
 \pdfoutput=1 
\usepackage{extsizes}
\usepackage[super,sort&compress,comma]{natbib} 
\usepackage[version=3]{mhchem}
\usepackage[left=1.5cm, right=1.5cm, top=1.785cm, bottom=2.0cm]{geometry}
\usepackage{balance}
\usepackage{mathptmx}
\usepackage{sectsty}
\usepackage{graphicx} 
\usepackage{lastpage}
\usepackage[format=plain,justification=justified,singlelinecheck=false,font={stretch=1.125,small,sf},labelfont=bf,labelsep=space]{caption}
\usepackage{float}
\usepackage{fancyhdr}
\usepackage{fnpos}
\usepackage[english]{babel}
\addto{\captionsenglish}{%
  \renewcommand{\refname}{Notes and references}
}
\usepackage{array}
\usepackage{droidsans}
\usepackage{charter}
\usepackage[T1]{fontenc}
\usepackage[usenames,dvipsnames]{xcolor}
\usepackage{setspace}
\usepackage[compact]{titlesec}
\usepackage{hyperref}
\usepackage{braket}
\usepackage{url}
\usepackage{color}
\usepackage{mathtools}
\usepackage{empheq} 

\definecolor{cream}{RGB}{222,217,201}

\begin{document}

\pagestyle{fancy}
\thispagestyle{plain}
\fancypagestyle{plain}{
\renewcommand{\headrulewidth}{0pt}
}

\makeFNbottom
\makeatletter
\renewcommand\LARGE{\@setfontsize\LARGE{15pt}{17}}
\renewcommand\Large{\@setfontsize\Large{12pt}{14}}
\renewcommand\large{\@setfontsize\large{10pt}{12}}
\renewcommand\footnotesize{\@setfontsize\footnotesize{7pt}{10}}
\makeatother

\renewcommand{\thefootnote}{\fnsymbol{footnote}}
\renewcommand\footnoterule{\vspace*{1pt}%
\color{cream}\hrule width 3.5in height 0.4pt \color{black}\vspace*{5pt}} 
\setcounter{secnumdepth}{5}

\makeatletter 
\renewcommand\@biblabel[1]{#1}            
\renewcommand\@makefntext[1]%
{\noindent\makebox[0pt][r]{\@thefnmark\,}#1}
\makeatother 
\renewcommand{\figurename}{\small{Fig.}~}
\sectionfont{\sffamily\Large}
\subsectionfont{\normalsize}
\subsubsectionfont{\bf}
\setstretch{1.125} 
\setlength{\skip\footins}{0.8cm}
\setlength{\footnotesep}{0.25cm}
\setlength{\jot}{10pt}
\titlespacing*{\section}{0pt}{4pt}{4pt}
\titlespacing*{\subsection}{0pt}{15pt}{1pt}

\fancyfoot{}
\fancyfoot[LO,RE]{\vspace{-7.1pt}\includegraphics[height=9pt]{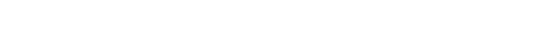}}
\fancyfoot[CO]{\vspace{-7.1pt}\hspace{13.2cm}\includegraphics{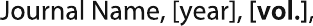}}
\fancyfoot[CE]{\vspace{-7.2pt}\hspace{-14.2cm}\includegraphics{head_foot/RF}}
\fancyfoot[RO]{\footnotesize{\sffamily{1--\pageref{LastPage} ~\textbar  \hspace{2pt}\thepage}}}
\fancyfoot[LE]{\footnotesize{\sffamily{\thepage~\textbar\hspace{3.45cm} 1--\pageref{LastPage}}}}
\fancyhead{}
\renewcommand{\headrulewidth}{0pt} 
\renewcommand{\footrulewidth}{0pt}
\setlength{\arrayrulewidth}{1pt}
\setlength{\columnsep}{6.5mm}
\setlength\bibsep{1pt}

\makeatletter 
\newlength{\figrulesep} 
\setlength{\figrulesep}{0.5\textfloatsep} 

\newcommand{\topfigrule}{\vspace*{-1pt}%
\noindent{\color{cream}\rule[-\figrulesep]{\columnwidth}{1.5pt}} }

\newcommand{\botfigrule}{\vspace*{-2pt}%
\noindent{\color{cream}\rule[\figrulesep]{\columnwidth}{1.5pt}} }

\newcommand{\dblfigrule}{\vspace*{-1pt}%
\noindent{\color{cream}\rule[-\figrulesep]{\textwidth}{1.5pt}} }

\makeatother

\newcommand*\mycommand[1]{\texttt{\emph{#1}}}
\newcommand{\yamazaki}[1]{\textcolor{cyan}{#1}}

\twocolumn[
  \begin{@twocolumnfalse}
{\includegraphics[height=30pt]{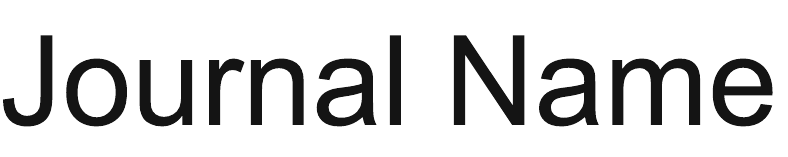}\hfill\raisebox{0pt}[0pt][0pt]{\includegraphics[height=55pt]{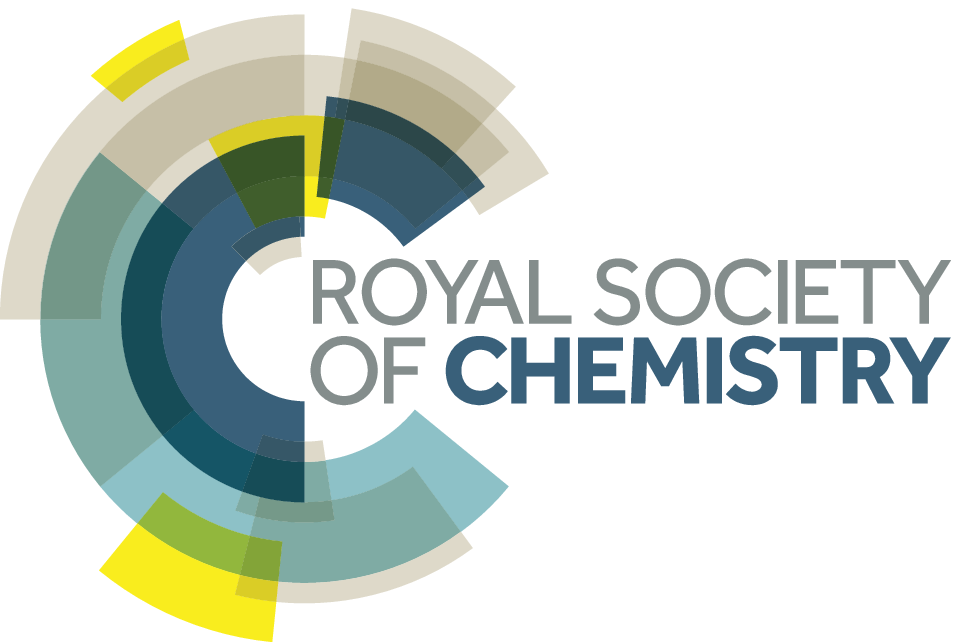}}\\[1ex]
\includegraphics[width=18.5cm]{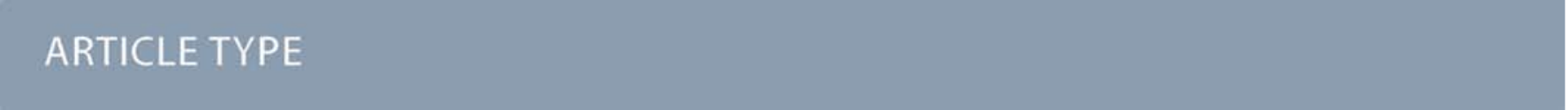}}\par
\vspace{1em}
\sffamily
\begin{tabular}{m{4.5cm} p{13.5cm} }

\includegraphics{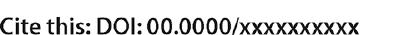} & \noindent\LARGE{\textbf{Population Trap in X-ray-induced Ultrafast Nonadiabatic Dynamics of Tropone Probed at the O(1\textit{s}) pre-edge$^\dag$}} \\
\vspace{0.3cm} & \vspace{0.3cm} \\

 & \noindent\large{Kaoru Yamazaki,$^{\ast}$\textit{$^{a}$} and Katsumi Midorikawa\textit{$^{a}$}} \\

\includegraphics{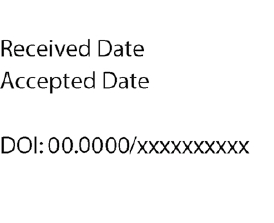} & \noindent\normalsize{Nonadiabatic transition (NAT) drives a variety of x-ray-induced photochemistry and photophysics used in nature and various fields. To clarify the x-ray-induced NAT dynamics, we performed nonadiabatic molecular dynamics simulations on electronically excited tropone (Tr) dications created by the carbon $KLL$ normal Auger decay. The Tr$^{2+}$ undergoes the NAT cascade via 10-10$^2$ states with time constants of 200-400 fs. We observed population traps in the highly excited states in 100 fs during the NAT cascade. The fingerprint of this population trap can be extracted from C($1s$) edge pump O($1s$) pre-edge probe femtosecond transient x-ray  absorption spectra  measured by the O($1s$) Auger electron yield method (TR-AEYS) using intense narrow band femtosecond x-ray free electron laser pulses. Our coupled ionization rate equation model demonstrates that selective and saturable C($1s$) core-ionization of Tr realizes background-free measurement. These results indicate that the importance of NAT in x-ray photochemistry and photophysics in large molecules. The real-time tracking of the NAT dynamics using TR-AEYS shall be a powerful approach for deeper insight.} \\

\end{tabular}

 \end{@twocolumnfalse} \vspace{0.6cm}
]


\renewcommand*\rmdefault{bch}\normalfont\upshape
\rmfamily
\section*{}
\vspace{-1cm}


\footnotetext{\textit{$^{a}$~Attosecond Science Research Team, Extreme Photonics Research Group, RIKEN Center for Advanced Photonics, RIKEN, 2-1 Hirosawa, Wako, Saitama, 351-0198, Japan. Fax: +81 48-462-4682; Tel: +81-50-3502-5574; E-mail: kaoru.yamazaki@riken.jp}}

\footnotetext{\dag~Electronic Supplementary Information (ESI) available: The details of computational methods, supplemental discussions, and parameters for LVC potentials. See DOI: 00.0000/00000000.}



\section{Introduction}
X-ray-induced photochemistry and photophysics are widely involved in nature and various fields . In biology, x-ray radiation triggers strand break of DNA double helix,resulting in cancer\cite{Sutherland2000,Azzam2012,Lomax2013,Alizadeh2015,Hans2021,Hahn2021}. In nanomedicine, x-ray deposited in nano-structures can be used for cancer treatments\cite{Wang2018_tps,Gong2021_ijnm,Geng2021_natchem,Guo2022_angewandte,Wang2022_natcomm}. 
In material science, organic scintillators convert x-ray radiation into low-energy light emission\cite{Wright1955,Buechele2015,Maddalena2019,Wang2021_natphoto,Gan2022,Wang2022_natcomm}. 

Such x-ray-induced processes are initiated by creating multiply charged cations populated in more than 10$^2$ or 10$^3$ electronically excited dicationic states via core ionization and subsequent Auger decay\cite{li2015_prl,Jahnke2021}. The excited dications undergo nonadiabatic transitions (NATs) to the lower-lying states via many intermediate states. The NATs after the Auger decay promote the ultrafast processes in x-ray photochemistry and photo-physics in small molecular systems such as sub 10 fs proton migration between the water dimer dication (\ce{(H2O)2^{2+}}) in liquid water \cite{Thurmer2013}, femtosecond C-C bond breaking in acetylene dication (\ce{C2H2^{2+}})\cite{li2017_natcomm}, and femtosecond vibronic charge migration in glycine dication (\ce{NH2CH2COOH^{2+}}) \cite{li2015_prl,Schwickert2022_sciadv}. Understanding the NAT dynamics after the Auger decay is key to clarifying the x-ray response of various molecular systems both in gas and condensed phases.

However, there is a debate on the contribution of the NATs in the electronically excited dications in the gas phase created by the x-ray irradiation for larger molecules.
Inhester et al. performed an x-ray photofragmentation experiment at the C($1s$) edge of ethyltrifluoroacetate (\ce{CF3COOC2H5}) using synchrotron-based photoelectron–photoion–photoion coincidence measurements and charge distribution analysis of the electronically excited \ce{CF3COOC2H5^{2+}} created by the carbon $KLL$ normal Auger decay using ab-initio electronic structure calculations\cite{inhester2018}.  They suggested that the fragmentation pattern is mostly determined by the initial hole distribution of  dications just after the Auger decay and the contribution of NAT is negligibly small\cite{inhester2018}.
 Kukk et al. conducted  the static and time-resolved Auger electron photo-ion-ion coincidence measurements using synchrotron and soft x-ray free electron laser (XFEL) facilities, respectively, and tight-binding reaction dynamics simulations, which are dedicated to the fragmentation dynamics of thiophene dication (\ce{C4H4S^{2+}}) created by the S(2\textit{p}) inner-shell ionization and subsequent Auger decay\cite{Kukk2015,kukk2021}. They argued  that the NAT from highly excited dicationic states to lower ones converts electronic energy to vibrational energy in the timescale 10$^2$ fs  and that the fragmentation occurs in the lower-lying states. This NAT could make the fragmentation statistical.

 To obtain a complete picture of the x-ray-induced NATs of large polyatomic molecules and deeper insights on the question above, ultrafast transient x-ray absorption spectra (TR-XAS) measurement using x-ray free electron lasers (XFELs)  \cite{Wolf2017natcomm,Lemke2017,Loh2020,Bergmann2021,Barillot2021,Schwickert2022_sciadv,Ross2022} or high-harmonic generation\cite{Loh2013_jpcl,Atter2017,Bhattacherjee2018_jacs,Bhattacherjee2018,Saito2019,Saito2021, Zinchenko2021,Scutelnic2021,Sidiropoulos2021,Garratt2022,Matselyukh2022} light sources is a prominent approach since they have realized the real-time tracking of the  UV/Vis and intense IR induced nuclear and valence electron dynamics with femto- to atto-second temporal resolutions in any form of matter\cite{Kraus2018_nrc,Neville2018_prl,Ehlert2018,Hua2019,Khalili2019,Picon2019,morzan2020,Frati2020,List2020,Khalili2020,Golubev2021,Shakya2021,Cistaro2021,Segatta2020,Segatta2021,Freibert2021,Soley2022}.
 Namely, TR-XAS measured by the Auger electron yield method (TR-AEYS), which uses the fact that the Auger electron yield at the probe process is proportional to the photoabsorption cross section, is becoming a useful tool with XFEL to investigate the x-ray-induced dynamics. \cite{Palacios2020,Driver2020,Barillot2021,Li2021_fd,Li2022_science,Khalili2021,Schwickert2022_sciadv}. 
 
To interpret the measured TR-XAS and TR-AEYS, the detailed theoretical investigations of the NAT dynamics and its associated TR-XAS/TR-AEYS are essential\cite{Loh2020,Zinchenko2021,Schwickert2022_sciadv}.  The combination of the ab-initio simulation of the Auger spectrum, which determines the initial states of NATs and nonadiabatic molecular dynamics method including the all two-hole dicationic states\cite{li2015_prl,li2017_natcomm} is a straight forward approach to simulate the NAT dynamics and associated associated TR-XAS/TR-AEYS. This approach has successfully applied to small molecules such as acetylene\cite{li2017_natcomm} and glycine \cite{li2015_prl} but not yet to the larger ones since the computational costs exponentially increase as a function of molecular size.

In this study, we theoretically investigated the NAT dynamics of an aromatic molecule tropone (\ce{C7H6O}, Tr)\cite{Ogasawara1972,hartman2017} before fragmentation triggered by the x-ray irradiation at the C($1s$) edge in the gas phase. We simulated the associated femtosecond C($1s$) edge pump  O($1s$) pre-edge probe femtosecond TR-AEYS. Tr is a good showcase of the x-ray-induced NAT dynamics since it contains a C=O group in its aromatic skeleton like the nucleobases and the other aromatic molecules used in organic x-ray scintillators\cite{Buechele2015,Maddalena2019,Wang2021_natphoto,Gan2022,Wang2022_natcomm}. The pre-edge absorption bands originated from the electronic transition from the 1\textit{s} core orbitals to valence holes are sensitive to the electronic structure of cationic systems \cite{Loh2020,Zinchenko2021,Barillot2021}.
The final states of the carbon $KLL$ Auger decay were determined by the two-hole population analysis technique, which calculates the relative intensity of the normal Auger spectra as a function of the electron population on the core-hole atoms\cite{mitani2003,inhester2018}. To simulate the subsequent NAT dynamics, we used a computationally efficient full-dimensional surface hopping nonadiabatic reaction dynamics technique on the pre-computed linear vibronic coupling model potential energy surface (LVC-MD)\cite{Plasser2018,heim2020,zobel2021}. We explicitly considered all valence two-hole states of \ce{Tr^{2+}} (210 singlet and 190 triplet states) via configuration interaction theory including all valence two-hole electron configurations \cite{li2015_prl}. We also investigated a possible measurement scheme for the TR-XAS measurement via coupled ionization rate equation model\cite{guichard2013}.

We found the population traps in highly excited dicationic states in 100 fs as experimentally observed in \ce{C4H4S^{2+}}\cite{kukk2021} during the cascade of NATs (NATs cascade) passing through 10-10$^2$ two-hole states. The time constants of NATs extracted from the calculated TR-AEYS clearly reflects the population trap.  Such x-ray induce NAT dynamics shall be extracted from the TR-AEYS. The TR-AEYS can be measured ideally in a background-free mode with an intense narrow band femtosecond XFEL light source. 

\section{Computational methods}

\subsection{Population dynamics of electronically excited \ce{Tr^{2+}}}

We consider the population dynamics on the $N$ dicationic states $\{\ket{\Psi_ \nu}\}$ $(\nu = 1, 2,\dots, N)$ of organic molecules created by the one-photon C($1s$) photoemission with an x-ray pulse and the subsequent normal Auger decay. The population of $\{\ket{\Psi_\nu}\}$ in incoherent limit $\mathbf{P} (t) = [P_1 (t), P_2(t),\dots, P_\nu(t),\dots, P_N(t)]^T$ is written as \cite{fukuzawa2019,heim2020},
\begin{equation}
    \mathbf{P} (t) = g(t,s_\text{pu}) \star \Theta(t)[1-\exp(-\tau_\text{Auger}^{-1}t)]\tilde{\mathbf{P}} (t),
\end{equation}
where $g(t,s_\text{pu})$ is the envelope function of the Gaussian pump pulse with the full width half maximum (FWHM) of $s_\text{pu}$, "$\star$'' represents convolution, $\Theta(t)$ is the Heaviside step function, $\tau_\text{Auger}$ is the overall Auger rate constant for the parent molecule, and  $\tilde{\mathbf{P}} (t) = [\tilde{P}_1 (t), \tilde{P}_2(t),\dots, \tilde{P}_\nu(t),\dots, \tilde{P}_N(t) ]^T$ is the population of the $\{\ket{\Psi_\nu}\}$ during the NAT dynamics in sudden ionization limit. The $\tilde{\mathbf{P}} (t)$ can be directly evaluated via nonadiabatic reaction dynamics simulations by setting the initial populations of the dicationic states of the neutral molecule $\{\tilde{P}_\nu(t=0)\}$ to be proportional to the relative Auger intensity $I_\nu$ ($\tilde{P}_\nu(t=0) = I_\nu / \sum_\nu I_\nu$) \cite{li2015_prl}.

For \ce{Tr^{2+}}, we considered  all $\{\ket{\Psi_\nu}\}$ ($N = 400$: 210 singlet and 190 triplet states) that has two holes in the 20 valence MOs generated by the carbon $KLL$ normal Auger decay. The relative Auger intensity was evaluated by the two-hole population analysis technique \cite{mitani2003,inhester2018}. We calculated the relative Auger transition amplitude from the C($1s$) core-hole MOs to valence two-hole electron configuration as a function of the electron population on the core-hole atoms according to Mitani et al.\cite{mitani2003}. The relative Auger intensity were evaluated by including the configuration interactions among the two-hole electron configurations and orbital relaxation effect \cite{inhester2018}.  The detailed formation can be found in  Electronic Supplementary Information (ESI). We adopted $\tau_\text{Auger} = 6.1$ fs according to an experimental value on benzene \cite{myrseth2002} and set $s_\text{pu} = 10 $ fs.

To evaluate the time evolution of $\tilde{\mathbf{P}} (t)$, we performed  full-dimensional surface hopping nonadiabatic reaction dynamics simulations using an Tully's fewest switching algorithm \cite{granucci2010,May2014,Mai2018,zobel2021}  
combined with the linear-vibronic coupling (LVC) model Hamiltonian\cite{Plasser2018,heim2020,zobel2021} (LVC-MD). 
We included both internal conversion and intersystem crossing. The LVC-MD initially constructs full dimensional harmonic oscillator potential energy surfaces 
for all electronic states of interest including vibronic and spin-orbit couplings\cite{Plasser2018}. Surface hopping simulations on the pre-computed LVC potential 
energy surfaces allow us to treat larger molecules with reasonable computational cost and statics \cite{heim2020,zobel2021} than 
conventional on-the-fly nonadiabatic molecular dynamics methods \cite{li2015_prl,li2017_natcomm}. Our approach cannot treat dissociation.
Thus, each trajectory was stopped if it reaches the electronic states where any bond-length elongates to the threshold bond-length $R_\text{th}=$ 3.4 \AA,
which corresponds to the double of the Van der Waals radius of carbon atom. After satifyin the stop criteria, we  thereafter regarded the \ce{Tr^{2+}} as "vibrationally hot" molecule and excluded it from the statistics. 

\subsection{TR-AEYS}

In this study, we focus on the O($1s$)$\rightarrow$valence hole transitions of \ce{Tr^{2+}} at O($1s$) pre-edge. Such "core to valence hole" transition is sensitive to the energy level of cationic states and the NAT dynamics among them\cite{Loh2020,Zinchenko2021}. \ce{Tr^{2+}} ejects an Auger electron via the resonant Auger decay after the core excitation. We then measure the yield of the Auger electron $N_\text{e}(q=2)$ as a function of probe photon energy $\hbar\omega_\text{pr}$ and pump-probe delay time $\Delta t$. We neglect the O($1s$)$\rightarrow$ virtual orbitals transitions, which will appear in the O($1s$) near-edge region. These transitions surely reflect the electronic structure of cations but their interpretation is difficult due to the spectral overlap among many transitions\cite{Loh2020,Zinchenko2021}. Furthermore, the simultaneous calculation of O($1s$)$\rightarrow$valence hole and O($1s$)$\rightarrow$virtual orbitals transitions for \ce{Tr^{2+}} in various electronically excited states is computationally very expensive.

The C($1s$) edge pump O($1s$) pre-edge probe femtosecond TR-AEYS of \ce{Tr^{2+}} $N_\text{e} (q=2, \hbar \omega_\text{pr}, \Delta t)$ can be calculated as a function of probe photon energy $\hbar \omega_\text{pr}$ and peak intensity of probe pulse $I_\text{pr}$ and pump-probe delay time $\Delta t$ defined by the time interval between the intensity peaks of the pump and probe pulse,
\begin{align}
    &N_\text{e} (q=2, \hbar \omega_\text{pr}, \Delta t) \notag\\ 
    &\propto \sum_{\nu, \mu} 
    g_\text{pr}(\hbar \omega_\text{pr}, \Delta t) \notag\\ 
    &\star
     P_{\nu}(\Delta t) 
    \sigma_{\nu \mu}  (\mathbf{R}(\Delta t)) 
    \delta( \hbar \omega_\text{pr} - \Delta E_{\nu \mu} (\mathbf{R}(\Delta t)), \Delta t)
    \frac{I_\text{pr}}{\hbar \omega_\text{pr}},
\end{align}
where $\mathbf{R}(\Delta t)$ is the molecular structure of the dication  at $\Delta t$, $\sigma_{\nu \mu} (\mathbf{R}(\Delta t))$ and $\Delta E_{\nu \mu} (\mathbf{R}(\Delta t))$ is the photoabsorption cross section and transition energy between the valence two-hole state $\ket{\Psi_{\nu} (\mathbf{R}(\Delta t))}$ and O($1s$) core-excited state $\ket{\Psi_{\mu}(\mathbf{R}(\Delta t))}$, respectively. The $\delta( \hbar \omega_\text{pr} - \Delta E_{\nu \mu} (\mathbf{R}(\Delta t)), \Delta t)$ is the two dimensional delta function located at $ \hbar \omega_\text{pr} = \Delta E_{\nu \mu} (\mathbf{R}(\Delta t)$ and $\Delta t$, and $ g_\text{pr}(\hbar \omega_\text{pr}, \Delta t)$ is the Gaussian instrumental function which determine the energy and temporal resolutions of the TR-AEYS.
To reduce the computational costs of the TR-AEYS calculation, we evaluated the $\ket{\Psi_{\nu} (\mathbf{R}(\Delta t)}$ and $\Delta E_{\nu \mu} (\mathbf{R}(\Delta t))$ in a diabatic representation characterized by the electron configurations at the Franck-Condon geometry  $\mathbf{R}_0$\cite{May2014,zobel2021}. We then applied the Condon approximation $\sigma_{\nu \mu}(\mathbf{R}(\Delta t))  = \sigma_{\nu \mu} (\mathbf{R}_0)$ .

\subsection{Quantum chemistry calculations}

The relative Auger intensity, the potential energy, vibronic \cite{Plasser2018} and spin-orbit couplings \cite{Hess1996,Malmqvist2002} among $\{\ket{\Psi_ \nu}\}$ appeared in the LVC model Hamiltonian were evaluated by the state-averaged complete active space multi-configuration self-consistent field (SA-CASSCF) theory\cite{Roos1980,aquilante2020} combined with the ANO-RCC-VTZP basis set \cite{Roos2004}. All twenty valence MOs and two holes (= 38 electrons) were included in the RAS2 active space (the [2h,20o] RAS2 space). All two-hole states resulting from the RAS2 space were included in the state-average procedure. We performed geometry optimization and normal mode analysis at the density fitted localized M\"oller-Plesset second-order perturbation theory (DF-LMP2)\cite{Werner2020} combined with the Def2-TZVP basis set\cite{Weigend2006} as implemented in the MOLPRO 2019.2 quantum chemistry package \cite{Werner2020}.   
The $\sigma_{\nu \mu} (\mathbf{R}_0)$ were calculated by the SA-2h-CASSCF/ANO-RCC-VTZP method by extending RAS2 space to 20 valence MOs and O(1$s$) MO, and by the restricted active space state interaction (RAS-SI) method\cite{Malmqvist1989}. 
To optimize the $\hbar \omega_\text{pu}$ and $\hbar \omega_\text{pr}$, we calculated the x-ray photoelectron spectra (XPS) of the lowest energy low spin states of Tr$^{q+} $ $(q = 0,1,2)$ at the C($1s$) and O($1s$) edges with low-lying shake-up states. We used multi-state restricted active space second-order perturbation theory (MS-RASPT2)\cite{Malmqvist2008,aquilante2020} combined with the Supporo-2012-TZP basis set for C and O atoms, and with the Supporo-2012-DZP basis set for H atoms \cite{Noro2012}. The relative intensity of the XPS was evaluated using the norm of Dyson orbitals. 
%
The OpenMolcas 18.09 and 21.10 quantum chemistry packages\cite{Galvan2019_jctc,aquilante2020} were used for all SA-CASSCF and MS-RASPT2 calculations, respectively. We used SHARC 2.1 program for the LVC-MD \cite{Mai2018}.
Further  details were described in ESI.

\section{Results and discussions}

\subsection{Structural and population dynamics during the NATs}

Let us look at the calculated NAT dynamics of \ce{Tr^{2+}}. Hereafter, the $S_n$ ($n = 0, 1,\dots , 209$) and $T_{n'}$ ($n' = 1, 2,\dots , 190$) denote the singlet and triplet diabatic states at the Franck-Condon geometry $\mathbf{R}_0$ of Tr, respectively. \ce{S}$_n$ and \ce{T}$_{n'}$ denote the singlet and triplet adiabatic states at a given geometry $\mathbf{R}(t)$ at time $t$, respectively.

\subsubsection{Carbon $KLL$ Auger spectrum}

The calculated Carbon $KLL$ normal Auger spectrum as a function of two-electron binding energy $E_\nu$  
is displayed in Figure 1a. The integrated intensity ratio of singlet and triplet states is singlet:triplet = 81:19 and Auger decay to singlet dicationic states is dominant. For the convenience of the later discussions, we divide the spectrum into three bands based on the peaks in the Auger spectrum and the character of each peaks.  Band I: $E_\nu <$ 30 eV, Band II: $E_\nu$ = 30-40 eV, and Band III: $E_\nu \ge$ 40 eV. According to the occupation numbers  of the natural orbitals $\{N_\text{occ}\}$ calculated at the [2h,20o]-SA-CASSCF/ANO-RCC-VTZP//DF-LMP2/Def2-TZVP level of theory, band I is predominantly generated by the low-lying two-hole dicationic states whose holes are located in the MOs originated form the 2$p$ atomic orbitals ($2p^{-2}$ states). The band II is the mixture of the high-lying $2p^{-2}$ and low-lying $2p^{-1}\sigma_{2s}^{-1}$ states.
For band III, the strongest peak in $E_\nu = 40-50$ eV originates from the high-lying $2p^{-1}\sigma_{2s}^{-1}$ states. The shoulder peaks in $E_\nu = 50-60$ eV  come from the high-lying $2p^{-1}\sigma_{2s}^{-1}$ and low-lying $\sigma_{2s}^{-2}$ states. The peaks above $E_\nu = 60$ eV are predominantly created by the high-lying $\sigma_{2s}^{-2}$ states.

The most intense transition at band II is assigned to be the $S_{36}$ state at $E_\nu = 32.95$ eV, whose normalized intensity is
$I_\nu = 0.026$, and that of the III is assigned to be the $S_{111}$ state) with  $E_\nu = 43.95$ eV and $I_\nu = 0.019$, respectively. 
The major electronic configurations of both $S_{36}$ and $S_{111}$ states are $\pi^{-1} \sigma_{2s}^{-1}$ according to the $\{N_\text{occ}\}$   as shown in Figures 1b and 1c, respectively. 

\begin{figure}[ht]
  \includegraphics[width=1.0\linewidth]{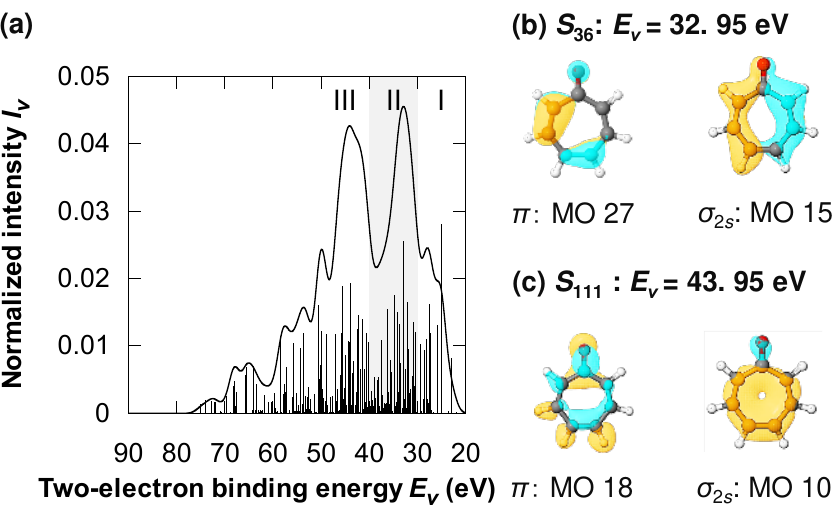}
  \caption{(a) The calculated normal Auger spectrum of tropone at carbon C($1s$) edge as a function of two-electron binding energy $E_\nu$. Peak position and normalized intensity ($\sum_\nu I_\nu  =1$) for each two hole states described with bar graph were calculated by the two-hole population analysis. The full curve represents the Gaussian convolution of FWHM = 2.5 eV. Band I: $E_\nu <$ 30 eV, Band II: $E_\nu$ = 30-40 eV, Band III: $E_\nu \ge$  40 eV. (b) and (c) The most intense transition for band II and III, respectively,  and the natural orbitals which has the two largest hole populations for each transition. The all results were calculated at the [2h,20o]-SA-CASSCF/ANO-RCC-VTZP//DF-LMP2/Def2-TZVP level of theory. }
  \label{fig:auger}
\end{figure}

\subsubsection{Population trap during the nonadiabatic transitions}

The NAT cascade proceeds mostly via internal conversion. The intersystem crossing is observed in only 16.5 \% of trajectories.
We observed population trap in bands III in 100-200 fs.
Figure 2a represents a representative trajectory of the NAT cascade. This trajectory started from the \ce{S117} state in band III relaxes into \ce{S4} in band I within 461 fs. \ce{Tr^{2+}} is trapped in band III until $t = 240$ fs since energy intervals of the dicationic states are so small that \ce{Tr^{2+}} can nonadiabatically return back to the higher-lying states as shown in Figure 2b. The mean interval of the singlet states in terms of $E_\nu$ was 0.15 eV in band III.  A large in-plane ring distortion observed at $t = 240$ fs (Figure 2a)  drives the internal conversion within $\sigma$ and $\pi$ orbitals, respectively. \ce{Tr^{2+}} slides down to the \ce{S4} state until by $t = 461$ fs and \ce{S2} state by $t = 476$ fs (the lowest lying adiabatic state in the present trajectory). The resultant vibrationally hot dications with $2p^{-2}$ electron configurations (I*) have a significant out-of-plane distortion around the C=O double bonds (Figure 2a, $t = 461$ fs), which promotes the hole transfer between $\sigma$  and $\pi$ orbitals as in the $^1\pi\pi^* \rightarrow  ~^1\text{n}\pi\*$ internal conversion in UV excited nucleobases and their derivatives\cite{Improta2016}. The dihedral angle between the C=O group and the two carbon atoms next to them reached 158 °. 

\begin{figure}[ht]
  \includegraphics[width=1.0\linewidth]{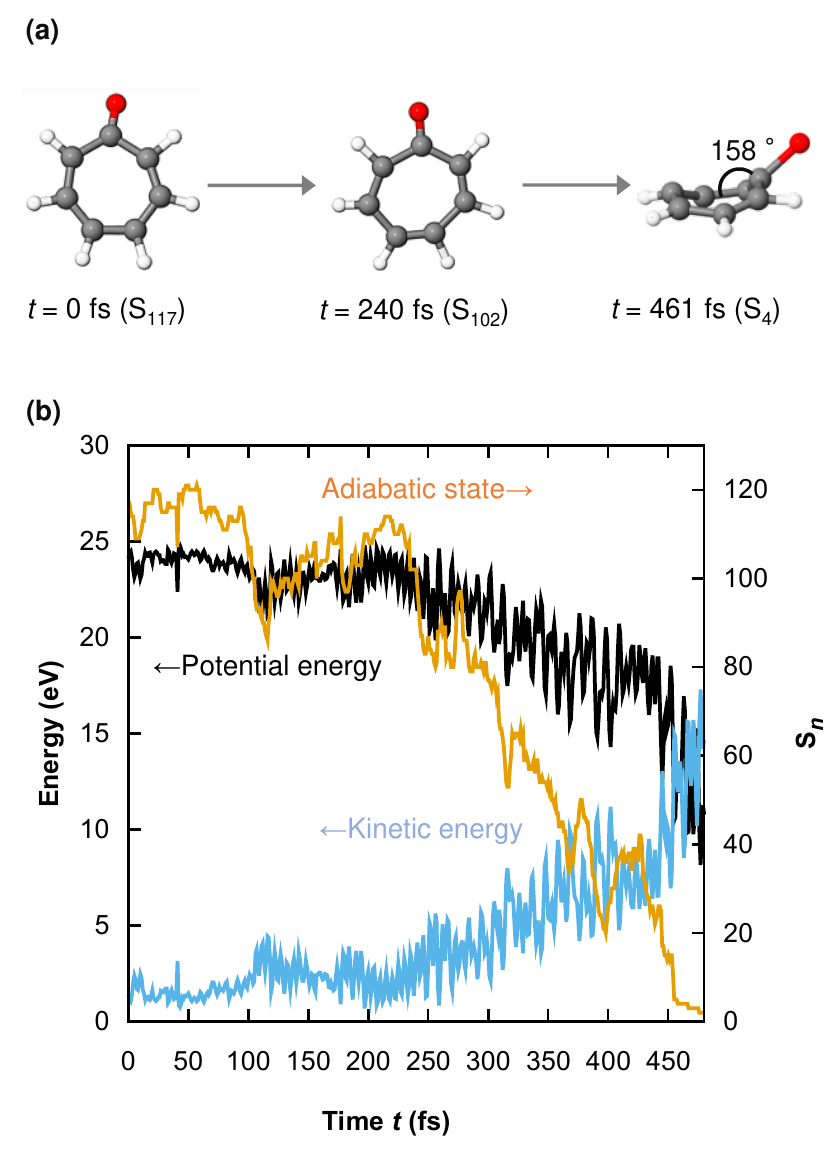}
  \caption{A representative trajectory of the NAT cascade of \ce{Tr^{2+}} starting from the band III calculated by the LVC-MD.
   (a) Selective snapshots of the trajectory and the adiabatic state S$_n$ is written in parenthesis.  (b) Potential energy (black, left axis ($\leftarrow$)), kinetic energy (blue, left axis ($\leftarrow$)), and adiabatic state S$_n$ (orange, right axis ($\rightarrow$)) as a function of time $t$. The origin of potential energy is set to be the $E_\nu$ of the ground (S$_0$) state of the dication (21.91 eV).  }
  \label{fig:trajectry}
\end{figure}


The $\mathbf{P}(t, E_\nu)$ clearly reflects the population trap. Figure 3a represents the  $\mathbf{P}(t, E_\nu)$ for \ce{Tr^{2+}} calculated by eq. 1, which is broadened along the $E_\nu$ axis with a Gaussian function of FWHM = 2.5 eV. Tr molecules were mainly ionized to  bands II and III after the carbon $KLL$ Auger decay in 20 fs. The population of the most prominent band III ($E_\nu = 44$ eV) is almost constant until $t = 100$ fs.  The electronically excited \ce{Tr^{2+}} in band III relaxed into band II within 200-400 fs. This refills the population in band II whose initial population relaxes to band I within 140 fs and makes the population of band II quasi-stational until $t = 200$ fs.  
\ce{Tr^{2+}} in band II further undergoes NAT to band I in 200 fs and completed the NAT cascade into the vibrationally hot lower-lying dicationic states in band I (band \ce{I^{*}}) in 100 fs.

\begin{figure}[ht]
  \includegraphics[width=1.0\linewidth]{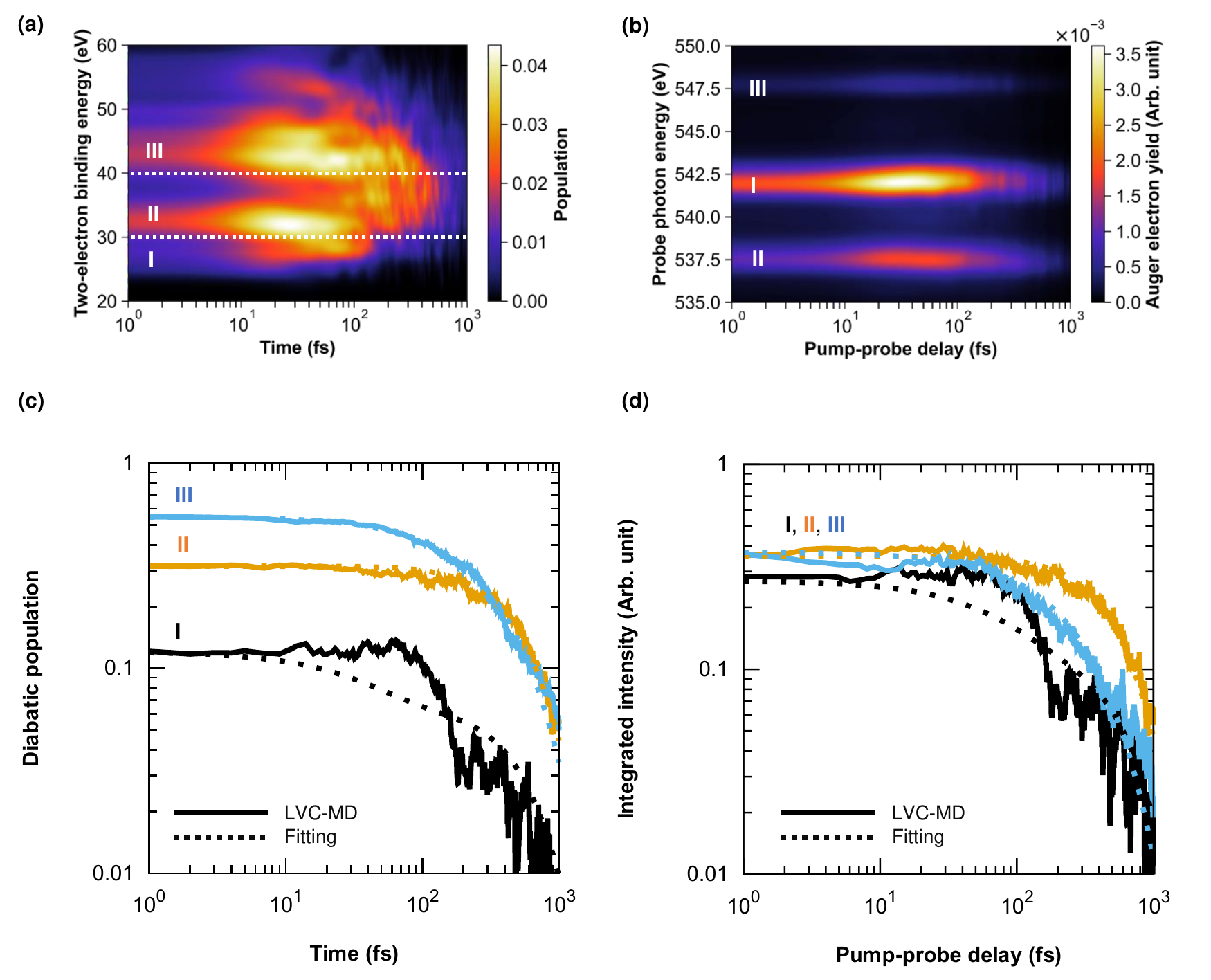}
  \caption{Population dynamics for the NAT cascade of \ce{Tr^{2+}} except vibrationally hot ones calculated by LVC-MD. 
  (a) The Diabatic population as a function of two-electron binding energy $E_\nu$ and time $t$ associated with the bands I-III. We assumed the FWHM of the Gaussian pump pulse is 10 fs and energy resolution in FWHM is 2.5 eV.
  (b) Calculated C($1s$) pump O($1s$) pre-edge probe TR-AEYS within the dipole and Condon approximation associated with the main contribution (band I, II, or III) for each absorption peak. We assumed that the FWHM of the Gaussian pump and probe pulses is 10 fs and energy resolution in FWHM is  1.0 eV.
  (c) Cumulative diabatic populations as a function of time and
  (d) Normalized absorption as a function of Pump probe delay for the band  I (Black), II (aqua), and III (orange) in sudden ionization limit. Solid lines represents the results from the LVC-MD, and dotted line represents fitting ones. We excluded vibrationally hot \ce{Tr^{2+}} and the integrated population becomes smaller than unity for panels (a) and (c). }
  \label{fig:pop_vs_trxas}
\end{figure}

To quantify the population dynamics associated with the population trap, we calculated the time-dependent diabatic population of \ce{Tr^{2+}} in sudden ionization limit in each bands $\tilde{p}_\chi (t) \equiv \sum_{E_\nu \in \chi}\tilde{P}_\nu(t) ~(\chi =  \text{III, II, I})$ from the LVC-MD results appearing as solid lines in Figure 3c. To extract the time constants $\{\tau\}$ for the NAT cascades which has sequential nature as disscused above, we constructed a kinetic model including the sequential NAT process of 
%
$
    \ce{III} \xrightarrow{\tau_{\mathrm{III}\rightarrow \mathrm{II}}}
    \ce{II} \xrightarrow{\tau_{\mathrm{II}\rightarrow \mathrm{I}}}
    \ce{I} \xrightarrow{\tau_{\mathrm{I}\rightarrow \mathrm{I}^*}}
    \mathrm{I}^*. 
$
  The coupled rate equations for this kinetic model can be written as
\begin{equation}
    \frac{\mathrm{d}}{\mathrm{d}t} \tilde{\mathbf{p}} (t)  =  \mathbf{k} \tilde{\mathbf{p}} (t),
\end{equation}
where $\tilde{\mathbf{p}} (t) \equiv \left[ \tilde{p}_\mathrm{III}(t), \tilde{p}_\mathrm{II}(t),\tilde{p}_\mathrm{I}(t)\right]^T$. The rate constant matrix $\mathbf{k}$ contains three time constants,
\begin{equation}
    \mathbf{k} 
    =  
    \begin{bmatrix}
       -\tau_{\mathrm{III}\rightarrow \mathrm{II}}^{-1} & 0 &0\\
       \tau_{\mathrm{III}\rightarrow \mathrm{II}}^{-1} & -\tau_{\mathrm{II}\rightarrow \mathrm{I}}^{-1} & 0 \\
       0 &\tau_{\mathrm{II}\rightarrow \mathrm{I}}^{-1} & -\tau_{\mathrm{I}\rightarrow \mathrm{I^*}}^{-1}  
    \end{bmatrix}.
\end{equation}
We evaluated the $\{\tau\}$ by a least square fitting into the $\tilde{\mathbf{p}} (t)$ calculated by the LVC-MD results. The initial value of the population $\tilde{\mathbf{p}} (t=0) $ were fixed to the LVC-MD values.

Fitted population transients are presented in Figure 3c. This panel  shows the temporal evolution of the three population
transients evaluated from the LVC-MD (solid lines) with the fitting functions (dotted lines); the kinetic model reasonably fits the data. The obtained time constants are shown in Table 1 together with their fitting errors. The resultant time constant of the NAT from band III to II $\tau_{\mathrm{III}\rightarrow \mathrm{II}}$  was  362 fs, which is longer than that of  band III to II ($\tau_{\mathrm{III}\rightarrow \mathrm{II}} = 211$ fs) due to the 100 fs population trap originating from the inverse internal conversion indicated in Figure 2b. After the NAT from II to I, the exited \ce{Tr^{2+}} in band I rapidly relaxes into I$^*$ ($\tau_{\mathrm{I}\rightarrow \mathrm{I}^*} = 41$ fs).

\begin{table}
  \caption{The optimized time constants $\{\tau\}$ associated with their standard fitting errors extracted for the NAT cascade using eqs 3 and 4. Comparison between the normalized  TR-AEYS signal $\Delta \tilde{\mathbf{N}}_\text{e} (\hbar \omega_\text{pr}, \Delta t)$ and diabatic population $\tilde{\mathbf{P}}(t)$. }
  \label{table:tau_trxas_vs_population}
   \centering
  \begin{tabular}{lrr}
    \hline
    Process   & Population (fs) & TR-AEYS (fs) \\
    \hline
    ${\mathrm{III}\rightarrow \mathrm{II}}$             & $362 \pm 1$ & $291 \pm 2$\\
    ${\mathrm{II}\rightarrow \mathrm{I}}$                 & $211 \pm 1 $& $274 \pm 1$\\
    ${\mathrm{I}\rightarrow \mathrm{I}^*}$                & $41 \pm 1 $& $84 \pm 1 $\\
    \hline
  \end{tabular}
\end{table}

\subsection{Fingerprints of the population trap in the TR-AEYS}

The time constants for the NAT dynamics can be extracted from TR-AEYS. Figure 3b shows the pre-edge peaks of
C($1s$) edge pump O($1s$) pre-edge probe femtosecond TR-AEYS. According to our MS-RASPT2 calculations, the vertical O($1s$) core ionization energy of Tr$^{2+}$, Tr$^{+}$, and
Tr at $\mathbf{R}_0$ are calculated to be 549.41, 542.58,  and 536.06 eV, respectively. 
The corresponding experimental value for Tr is 535.04 $\pm$ 0.2 eV \cite{Zhdanov1977}. 
For \ce{Tr^{2+}} (Figure 3(b)), there are three peaks at the probe photon energies of $\hbar \omega_\text{pr} = $548, 542, and 538 eV. Each peak consists of $\sim$10-10$^2$ core-to-valence
transitions $\{\nu \rightarrow \mu \}$ from various valence two-hole states $\{ \ket{\Psi_\nu} \}$. However, the peak at $\hbar \omega_\text{pr} = 548, 542, $ and 538 eV have the largest contributions from the transitions to bands III, I, and II, respectively. The intensity of the peak at the $\hbar \omega_\text{pr} =548$ eV is almost constant until $\Delta t =100$ fs as observed in the population of the most prominent peak of band III ($E_\nu = 44$ eV, Figure 3a) reflecting the population trap.  One electron picture conserves in the intense transitions from bands I and II. 
On the other hand, the weak transitions from band III have multi-electron nature and we observed not only the "core to valence hole" transitions but also associated shake-up and shake-off processes among the valence orbitals: another electron in valence MOs moves to the remaining vacancy which is not filled by the excited electron from the O($1s$) orbital. This makes the transition energy of the band III transition higher than those of bands I and II.

To extract the $\{\tau\}$ from the TR-AEYS of the O($1s$)-pre-edge region ($\hbar \omega_\text{pr} \le 550$ eV) of \ce{Tr^{2+}}, we first calculated the integrated absorption intensities originated from the transition from bands I-III in the sudden ionization and impulsive excitation limit defined as
$
\tilde{N}_{\text{e}\chi} (\Delta t)  
    \equiv
    \sum_{E_\nu \in \chi}\sum_{\mu} 
    \sigma_{\nu \mu}  (\mathbf{R}(\Delta t)) 
    \tilde{P}_{\nu}(\Delta t) 
    $ $/ 
    \sum_{\nu, \mu} 
    \sigma_{\nu \mu}  (\mathbf{R}(\Delta t)) 
    \tilde{P}_{\nu}(\Delta t)
$
from the LVC-MD results. We omit $q=2$ for clarity in this subsection. The resultant $\tilde{N}_{\text{e},\chi} (\Delta t) $ is shown in Figure 3d as the solid lines. The $\{\tau\}$ were extracted by using 
the same sequential NAT kinetic model as the analysis of  $\tilde{P}_\chi (\Delta t)$. The calculated integrated absorption intensities
$
\tilde{\mathbf{N}}_\text{e} (\Delta t) 
\equiv 
[ \tilde{N}_{\mathrm{e,III}}(\Delta t), 
  \tilde{N}_\mathrm{e,II}(\Delta t),
  \tilde{N}_\mathrm{e,I}(\Delta t)]^T
$ 
were fitted by the 
analytical solution of the coupled differential equation $\mathrm{d}\tilde{\mathbf{N}}_\text{e} (\Delta t)/\mathrm{d} \Delta t = \mathbf{k}\tilde{\mathbf{N}}_\text{e}(\Delta t)$. 
The initial value $\tilde{\mathbf{N}}_\text{e} (t = 0)$ was fixed to the LVC-MD values.

The fitting results on the TR-AEYS transients are presented in Figures 3d. The kinetic 
model (dotted lines) reasonably fits the LVC-MD results (solid lines). The extracted time constants $\{ \tau\}$ are summarized in Table 1 together with their fitting errors. The $\{ \tau\}$  extracted from $\Delta \tilde{N}_\text{e} (\hbar \omega_\text{pr}, \Delta t)$ agree with the reference $\{ \tau\}$ obtained from $\tilde{\mathbf{P}}(t)$: (i) the rate limiting step is the NAT from band III to II due to the population trap; (ii) The intraband NAT from band I to I$^*$ has the timescales of sub 100 fs.
TR-AEYS emphasizes the decay profiles between specific dicationic states which have optically allowed core-to-valence transitions, especially if one evaluated the averaged time constants using the integrated intensity over each band as in this study. 
This slightly deviates the TR-AEYS based $\{\tau\}$ to the population-based ones.

\subsection{Proposed experimental setup}

Let us discuss how to efficiently measure the TR-AEYS  \cite{Palacios2020,Driver2020,Barillot2021,Li2021_fd,Li2022_science,Khalili2021,Schwickert2022_sciadv} of Tr using coupled ionization rate equation model\cite{guichard2013}.
We found that a narrow band pump pulse whose photon energy $\hbar\omega_\text{pu}$ is set to be 292.3 eV to ionize only the C($1s$) orbitals of the neutral Tr and whose peak intensity $I_\text{pu}$ should be larger than the value of 10$^{15}$ W/cm$^2$ to saturate the ionization
of the neutral. For the probe pulse, its photon energy $\hbar \omega_\text{pr} $ should be set to the O($1s$) pre-edge region of Tr$^{2+}$ and $I_\text{pr} $ should be in the linear absorption regime ($\sim 10^{14}$ W/cm$^2$). Here, we first explain the analytical formula for the ion yields [Tr$^{q+}$] ($q = 0,1,2,3$) and the Auger electron yield $N_\text{e}$ in the pump-probe processes.  We included one- and two-photon sequential ionization processes by the pump pulse and one-photon ones by the probe pulse. Then, we discuss the optimal laser parameters for the proposed experiment.
%

The pump pulse can trigger the carbon $KLL$ normal Auger decay for Tr ($C$), valence ionization ($V^{-1}$) for Tr$^{q+} (q = 0,1,2)$, and core excitation from C$(1s)$ edge to valence holes and subsequent resonant Auger decay (C($1s$)$ \rightarrow V$) for Tr$^{+}$. The possible two ionization pathways by the pump pulse are 
$
    \ce{Tr}  \xrightarrow{C} \ce{Tr}^{2+} \xrightarrow{V^{-1} \text{ or } \text{C($1s$)} \rightarrow V} \ce{Tr}^{3+} 
$
and
$
    \ce{Tr}  \xrightarrow{V^{-1}} \ce{Tr}^{+} \xrightarrow{V^{-1} \text{ or } \text{C($1s$)} \rightarrow V} \ce{Tr}^{2'+} 
$
, where \ce{Tr^{2+}} is the dication of Tr which has two holes in valence MOs created by 
the $C$ process as discussed above and \ce{Tr^{2'+}} represents those created by the two-photon sequential double ionization process. 
The $C$ process from Tr to Tr$^{2+}$ contributes to the signal and the others do to the background. As the photo-ionization and absorption cross sections are almost independent of $q$ \cite{guichard2013}, the yield of Tr$^{2+}$ after the Gaussian  pump pulse irradiation whose peak intensity is $I_\text{pu}$ and full width at half maximum (FWHM) is $s_\text{pu}$ can be written as functions of $I_\text{pu}$ and $\hbar \omega_\text{pu}$ as
 \begin{align}
\label{eq_q2_pu_text}
    [\ce{Tr}^{2+}] (I_\text{pu}, \hbar \omega_\text{pu}) 
    &= [
        \exp \left( -\lambda_V(I_\text{pu}, \hbar \omega_\text{pu})  \right) \notag\\
        &-\exp\left( -\lambda_V(I_\text{pu}, \hbar \omega_\text{pu}) - \lambda_C(I_\text{pu}, \hbar \omega_\text{pu}) \right)],\\
%
    \label{eq:def_lambda_pu}
    \lambda_{A_\text{pu}}(I_\text{pu}, \hbar \omega_\text{pu}) 
    &= \sigma_{A_\text{pu}}(\hbar \omega_\text{pu}) \frac{I_\text{pu}}{\hbar \omega_\text{pu}}
    \sqrt{
    \frac{\pi}{4\ln2}
    } s_\text{pu} \notag\\ & (A_\text{pu} = V^{-1}, C, X\rightarrow V),
\end{align}
where $\sigma_{A_\text{pu}}(\hbar \omega_\text{pu}) $ represents the photoionization cross section of the ionization processes of $A_\text{pu}~(A_\text{pu} = V^{-1}, C, \text{C}(1s)\rightarrow V)$ as functions of $\hbar \omega_\text{pu}$, respectively. $\lambda_{V}(I_\text{pu},\hbar \omega_\text{pu})$ and $\lambda_{C}(I_\text{pu}, \hbar \omega_\text{pu})$ mean the number of absorbed photons during the pump irradiation for the ionization processes $V^{-1}$ and $C$, respectively. The analytical expressions for $[\ce{Tr}^{q+}](I_\text{pu}, \hbar \omega_\text{pu}) ~ (q=0,1,2')$ can be found in ESI.


One photon absorption from the probe pulse at the O($1s$)-pre-edge of Tr$^{2+}$ excites core electron at O($1s$)-edge to the valence hole. The core-excited Tr$^{2+}$ which is rapidly ionized into Tr$^{3+}$ via resonant Auger relaxation emits one fast Auger electron e$^{-}_\text{Auger}$,
$
    \ce{Tr}^{2+}  \xrightarrow{\ce{O}(1s) \rightarrow V} \ce{Tr}^{3+} + \ce{e}^{-}_\text{Auger}. 
$
The TR-AEYS measurement selectively counts up the number of e$^{-}_\text{Auger}$ as a function of $\hbar \omega_\text{pr}$ and $\Delta t$ by using electron spectrometers \cite{Palacios2020,Driver2020,Barillot2021,Li2021_fd,Li2022_science,Khalili2021,Schwickert2022_sciadv}. This scheme can eliminate photoelectrons from core ionization and $V^{-1}$ processes, and Auger electrons from the $C$ process of Tr$^{q+} (q = 0-3)$ by the pump and probe pulses. The major background sources can come from the resonant Auger electrons from the O$(1s) \rightarrow V$  excitation process of Tr$^+$ and Tr$^{2+}$, and normal Auger electron from O$(1s)^{-1}$ core ionization of Tr, which can overlap with the signal from \ce{Tr^{2+}}.

When the temporal overlap between the pump and probe pulse is negligibly small, the Auger electron yield $N_\text{e} $ originated from Tr$^{q+} (q = 0,1,2,2')$ can be written as,
\begin{align}
    \label{eq:electron_yield_text}
    &N_\text{e}(q, I_\text{pu}, \hbar \omega_\text{pu},I_\text{pr}, \hbar \omega_\text{pr}) \notag\\
    &= \theta (q, \hbar \omega_\text{pr})[\ce{Tr}^{q+}](I_\text{pu}, \hbar \omega_\text{pu})\frac{\sigma_{A_\text{pr}}(q, \hbar \omega_\text{pr})}{\sigma_\text{tot}(q,\hbar \omega_\text{pr})} \notag\\
    & \times \left[
        1-\exp(-\lambda_\text{tot}(q, I_\text{pr}, \hbar \omega_\text{pr}))
    \right]
\end{align}
Here, $\theta (q, \hbar \omega_\text{pr})$ describes the fraction of bright states for Tr$^{q+}$ at $\hbar \omega_\text{pr}$ ($\theta (q= 0, \hbar \omega_\text{pr}) = 1$ and $0\le \theta (0 < q, \hbar \omega_\text{pr}) \le 1$).  $\sigma_{A_\text{pr}}(q, \hbar \omega_\text{pr})$ represents the mean photoabsorption crossection for the O$(1s) \rightarrow V$ excitation and subsequent resonant Auger decay  for $q=1, 2,2'$, and the O$(1s)^{-1}$ core ionization and subsequent oxygen $KLL$ Auger decay ($O$) for $q=0$, respectively.
$\sigma_\text{tot}(q,\hbar \omega_\text{pr}) \equiv \sigma_{A_\text{pr}}(q, \hbar \omega_\text{pr}) + \sigma_{V}(\hbar \omega_\text{pr})+ \sigma_{C}(\hbar \omega_\text{pr})$  is the total photoionization cross section for Tr$^{q+}$ as a function of $\hbar \omega_\text{pr}$, and $\lambda_\text{tot}(q, I_\text{pr}, \hbar \omega_\text{pr})$ is the total number of absorbed photon in the probe irradiation. 
For the Gaussian probe pulse with a peak intensity of $I_\text{pr}$  and FWHM of $s_\text{pr}$, $\lambda_\text{tot}(q, I_\text{pr}, \hbar \omega_\text{pr})$ is written as
%
$
    \lambda_\text{tot}(q, I_\text{pr}, \hbar \omega_\text{pr}) 
    = \sigma_\text{tot}(q, \hbar \omega_\text{pr}) (I_\text{pr}/\hbar \omega_\text{pr})
        ({\pi}/{4\ln2})^{1/2} s_\text{pr}.
$
%
In the liner probe regime, eq \eqref{eq:electron_yield_text} is finally simplified to
\begin{align}
        \label{eq:electron_yield_liner_text}
    &N_\text{e}(q, I_\text{pu}, \hbar \omega_\text{pu},I_\text{pr}, \hbar \omega_\text{pr}) 
    \approx 
    \theta (q, \hbar \omega_\text{pr})
    [\ce{Tr}^{q+}](I_\text{pu}, \hbar \omega_\text{pu}) \notag\\
    & \times \sigma_{A_\text{pr}}(q, \hbar \omega_\text{pr})
    \frac{I_\text{pr}}{\hbar \omega_\text{pr}}
    \sqrt{
    \frac{\pi}{4\ln2}
    } s_\text{pr}.
\end{align}


The S/B $\eta$ can be defined as the ratio of $ N_\text{e}(q=2, I_\text{pu}, \hbar \omega_\text{pu},I_\text{pr}, \hbar \omega_\text{pr})$ and the sum of the electron yields originated from the other O$(1s)$-edge (resonant) Auger processes,
\begin{equation}
\label{eq:S/B_def_text}
    \eta( I_\text{pu}, \hbar \omega_\text{pu},I_\text{pr}, \hbar \omega_\text{pr}) 
    \equiv \frac{
    N_\text{e}(q=2, I_\text{pu}, \hbar \omega_\text{pu},I_\text{pr}, \hbar \omega_\text{pr})}{
    \sum_{q=0,1,2'}N_\text{e}(q, I_\text{pu}, \hbar \omega_\text{pu},I_\text{pr}, \hbar \omega_\text{pr})}.
\end{equation}

We have searched for the condition that the 
$N_\mathrm{e}(q, I_\mathrm{Ipu}, \hbar\omega_\mathrm{pu}, I_\mathrm{pr},\hbar\omega_\mathrm{pr})$ in eq. (7) and $\eta(I_\mathrm{Ipu}, \hbar\omega_\mathrm{pu}, I_\mathrm{pr},\hbar\omega_\mathrm{pr})$ in eq. (11) 
should be simultaneously maximized
upon scanning $I_\mathrm{pu}$ and $I_\mathrm{pr}$ at fixed $q$, $\hbar\omega_\mathrm{pu}$, and $\hbar\omega_\mathrm{pr}$.
To optimize $\hbar\omega_\text{pu}$, We calculated C($1s$) XPS spectra for \ce{Tr^{$q+$}} ($q = 0,1,2$) at the MS-RASPT2 level of theory are shown in Figures 4 and S2, respectively. Both C($1s$) and O($1s$) edges blue-shift by 6-7 eV for each valence ionization of $q \rightarrow q+1$. 
We have chosen $\hbar\omega_\mathrm{pu} = 292.3$ eV
to avoid sequential C($1s$) core ionization process in Tr$^{+}$ and Tr$^{2+}$ cations. We set $\hbar \omega_\text{pr} = 542.0$ eV
since the strongest O($1s$)$\rightarrow V$ absorption band of Tr$^{2+}$ appears here as shown in Figure 3b.
The charge number $q$ is set to 2. 
We defined a merit function 
$\Upsilon(I_\mathrm{pu}, I_\mathrm{pr}) \equiv N_\mathrm{e}(I_\mathrm{pu}, I_\mathrm{pr})\eta (I_\mathrm{pu}, I_\mathrm{pr})$
to be maximized,
where we omit the variables other than scanning variables for simplicity.
The scanning range of $I_\mathrm{pu}$ is determined to be
$I_\mathrm{pu}\ge \mathrm{1.0\times10^{15}~W/cm^{2} }$ to saturate the absorption of the pump pulse,
while that of $I_\mathrm{pr}$ is  determined to be
$ I_\mathrm{pr}\le \mathrm{1.0\times10^{14}~W/cm^{2}}$ to avoid nonlinear absorption of the probe pulse.
The values of cross sections\cite{Yeh1985} are summarized in Table S2 in ESI.

\begin{figure}[ht]
  \includegraphics[width=1.0\linewidth]{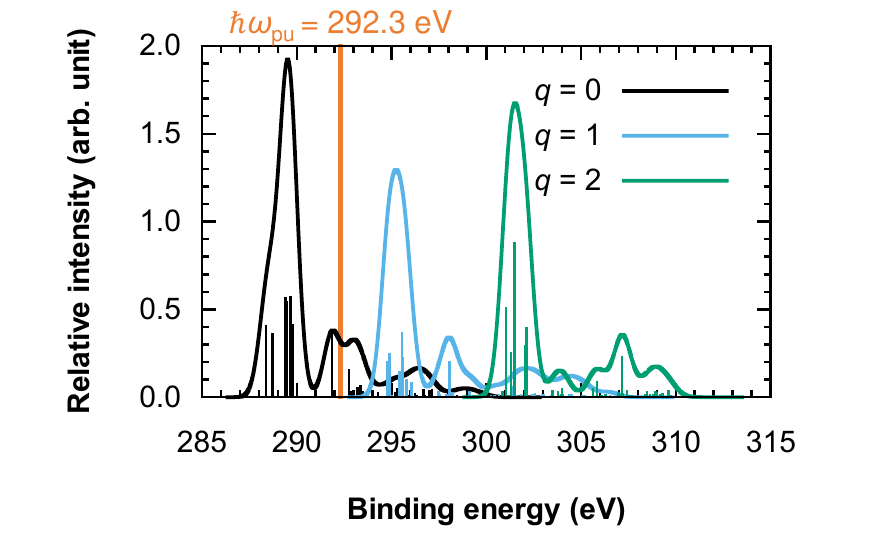}
  \caption{   
The calculated XPS spectra of Tr$^{q+}$ $(q = 0,1,2)$ at the C($1s$) edge. Results at the MS-RASPT2/Sapporo-2012 (C,O: TZP, H: DZP) level of theory. The calculated spectra were broadened by a Gaussian function of FWHM  = 1.0 eV. The position of $\hbar \omega_\text{pu}$ is shown in orange.
   }
\label{fig:c1sxps}
\end{figure}

We found that $\Upsilon(I_\text{pu},I_\text{pr})$ becomes maximized to $3.5 \times 10^{-2}$ at $I_\text{pu} = I_\text{opt,pu} = 3.7 \times 10^{15}$ W/cm$^2$ and $I_\text{pr} = I_\text{opt,pr} = 1.0 \times 10^{14}$ W/cm$^2$ as shown in Figure 5a. This optimized $I_\text{opt,pu}$ is a bit larger than to the saturable intensity $I_\text{max} = 2.0 \times 10^{15}$ W/cm$^2$ which maximizes $[\ce{Tr}^{2+}](I_\text{max})$ to 0.76 (Figure 5b) and $N_\text{e}(q=2, I_\text{max},I_\text{pu})$ to $N_\text{max}= 4.2 \times 10^{-3}$ to minimize the  background from Tr$^{2+}$.

At the optimized $I_\text{opt,pu}$ and $I_\text{opt,pr}$, $\eta(I_\text{opt, pu},I_\text{opt, pr})$  reaches 9.3 (Figure 5c), which is not sensitive to the value of $I_\text{pu}$ in the saturable regime ($\eta \ge 8.0$ for $I_\text{pu} = 3.0-6.5 \times 10^{15} $ W/cm${^2}$) and constant for the whole range of the investigated $I_\text{pr}$. The maximization of [Tr$^{2+}$] by the neutral selective core-ionization with a narrow band pump pulse (ideally a seeded x-ray free electron laser pulse\cite{Amann2012}) is the key to realizing the precise TR-AEYS measurement. The $N_\text{e}(q=2)=3.7 \times 10^{-3}$ (Figure 5d) indicates that a repetition rate of $\sim$0.1-1.0 MHz is essential for the data acquisition in a reasonable machine time. The suggested laser parameters for the pump and probe pulses and detection scheme have already been available at the NAMASTE end station of LCLS-II\cite{Walter2022}. 

\begin{figure}[ht]
  \includegraphics[width=1.0\linewidth]{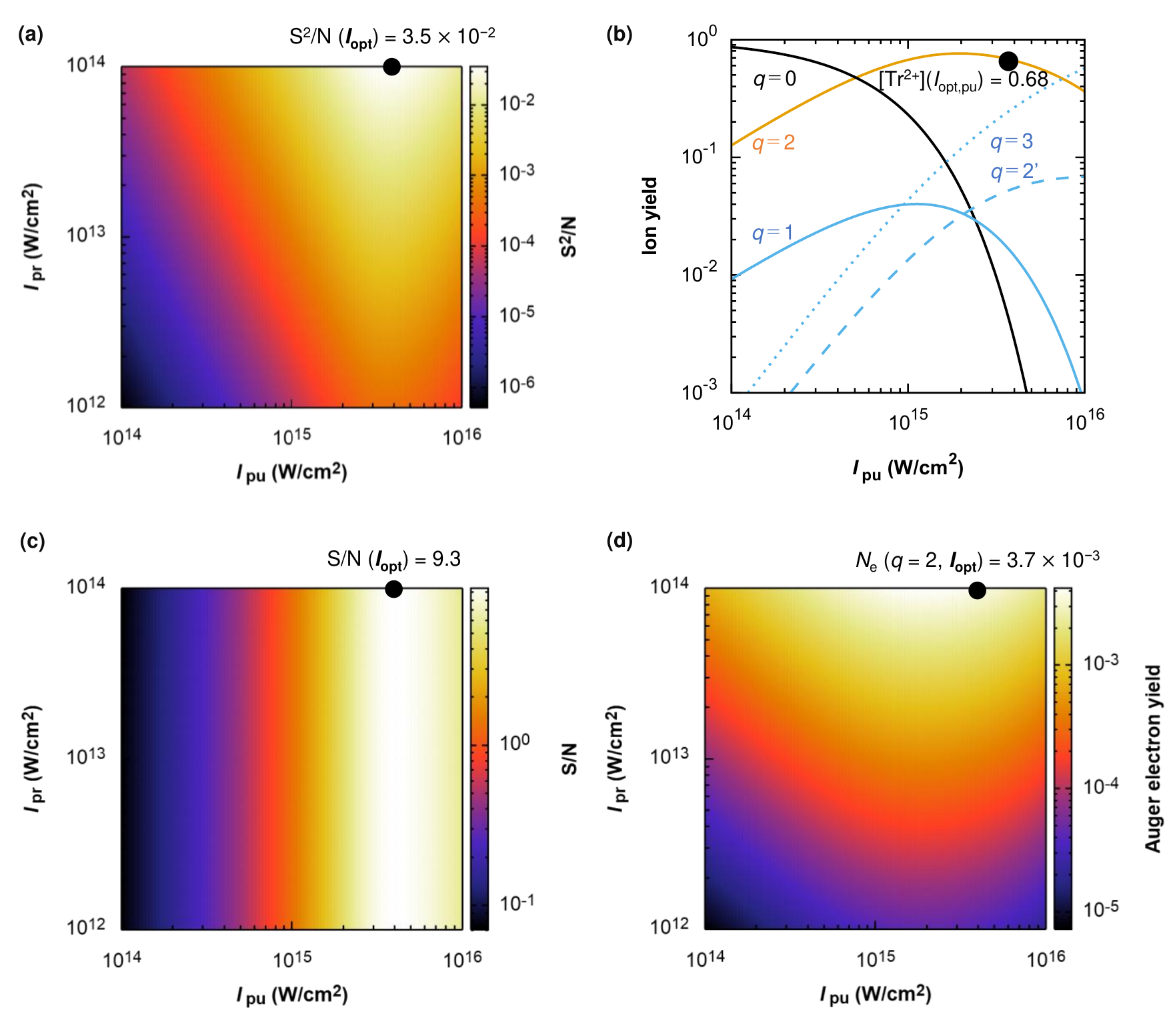}
  \caption{   
  (a) The S$^2$/N ratio $\Upsilon(I_\text{pu},I_\text{pr})$ at $\hbar \omega_\text{pu} = 292.3$ eV and $\hbar \omega_\text{pr} = 542.0$ eV.
  (b) The ion yields [Tr$^{q+}$]$(I_\text{pu})$ $(q = 0,1,2,2',3)$ at $\hbar \omega_\text{pu} = 292.3$ eV.
   (c) The S/B ratio $\eta(I_\text{pu},I_\text{pr})$ at $\hbar \omega_\text{pu} = 292.3$ eV and $\omega_\text{pr} = 542.0$ eV. We set $\theta(q= 0,\hbar \omega_\text{pr}) = 1.0$ and $\theta(q= 1,2,2',\hbar \omega_\text{pr}) = 0.10$.
  (d) The resonant Auger electron yield for \ce{Tr^{2+}} $N_\text{e} (q = 2,I_\text{pu},I_\text{pr})$ at $\hbar \omega_\text{pu} = 292.3$ eV and $\omega_\text{pr} = 542.0$ eV. 
   }
\label{fig:ionization}
\end{figure}

\section{Conclusions}
In conclusion, we simulated the carbon $KLL$ Auger induced nonadiabatic reaction dynamics of Tr that were revealed with C($1s$) edge pump  and O($1s$) pre-edge probe femtosecond TR-AEYS using LVC-MD and the coupled rate equation model. 
We found population traps in highly excited dicationic states in 100 fs as experimentally observed in \ce{C4H4S^{2+}} during the NAT cascade passing through 10-10$^2$ two-hole states. The time constants of NATs extracted from the calculated TR-AEYS reflect the population trap.  Such x-ray induce NAT dynamics shall be detected by the TR-AEYS. The TR-AEYS can be measured ideally in a background-free mode with an intense narrow band femtosecond XFEL light source.
For this background-free TR-ARES measurement,  a narrowband pump pulse whose photon energy is set to ionize only the C($1s$) MO of Tr and whose peak intensity is adjusted to be large enough to saturate the  C($1s$) core ionization. Our theoretical study demonstrates that it is essential to seriously consider the effect of NAT on the x-ray photochemistry and photophysics of large molecules. We also hope that the proposed two-color TR-AEYS scheme will by widely used for investigating the nonadiabatic nature of x-ray photochemisitry and photophysics.

\section*{Author Contributions}
\begin{description}
\item[K. Y.:]  Conceptualization, Methodology, Resources, Investigation, Formal analysis, Writing - Original Draft, and Funding acquisition. 
\item[K. M.:]  Resources,  Writing - Review \& Editing, Funding acquisition, and Supervision.    
\end{description}

\section*{Conflicts of interest}
There are no conflicts to declare.

\section*{Acknowledgements}
K.Y. acknowledges to JST-PRESTO (JPMJPR210A) for financial support. K. Y. and K. M. are grateful to the financial support
from JSPS KAKENHI Grant Number 19H05628. We also thank Drs. Tomoya Okino and Yasuo Nabekawa at RIKEN for their fruitful comments and discussions.  A part of calculations was performed using Research Center for Computational Science, Okazaki, Japan (Project numbers: 20-IMS-C505, 21-IMS-C502, and 22-IMS-C508).  



\balance

\renewcommand\refname{References}

\bibliography{library} 
\bibliographystyle{rsc} 

\begin{figure}[ht]
  \includegraphics[width=1.0\linewidth]{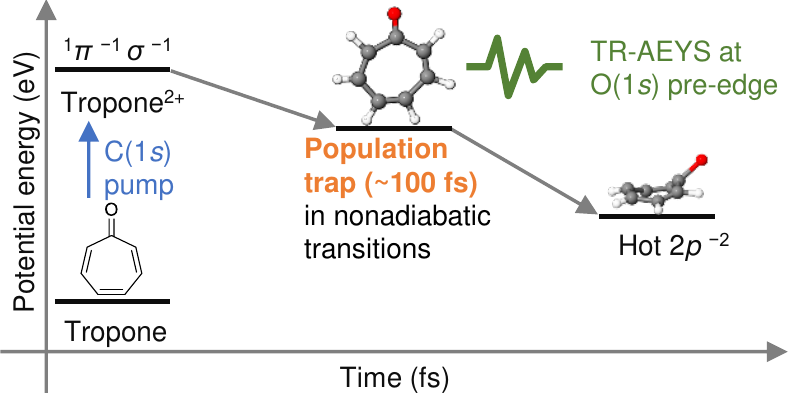}
  \caption*{   
 TOC figure: \ce{Tropone^{2+}} created by the carbon $KLL$ Auger decay are trapped in the highly excited electronic states in 100 fs during the nonadiabatic transitions. This can be traced by measuring transient Auger electron yield spectra (TR-AEYS) at the O($1s$) pre-edge.
   }
\label{fig:toc}
\end{figure}

\end{document}


\pagestyle{fancy}
\thispagestyle{plain}
\fancypagestyle{plain}{
\renewcommand{\headrulewidth}{0pt}
}

\makeFNbottom
\makeatletter
\renewcommand\LARGE{\@setfontsize\LARGE{15pt}{17}}
\renewcommand\Large{\@setfontsize\Large{12pt}{14}}
\renewcommand\large{\@setfontsize\large{10pt}{12}}
\renewcommand\footnotesize{\@setfontsize\footnotesize{7pt}{10}}
\makeatother

\renewcommand{\thefootnote}{\fnsymbol{footnote}}
\renewcommand\footnoterule{\vspace*{1pt}%
\color{cream}\hrule width 3.5in height 0.4pt \color{black}\vspace*{5pt}} 
\setcounter{secnumdepth}{5}

\makeatletter 
\renewcommand\@biblabel[1]{#1}            
\renewcommand\@makefntext[1]%
{\noindent\makebox[0pt][r]{\@thefnmark\,}#1}
\makeatother 
\renewcommand{\figurename}{\small{Fig.}~}
\sectionfont{\sffamily\Large}
\subsectionfont{\normalsize}
\subsubsectionfont{\bf}
\setstretch{1.125} 
\setlength{\skip\footins}{0.8cm}
\setlength{\footnotesep}{0.25cm}
\setlength{\jot}{10pt}
\titlespacing*{\section}{0pt}{4pt}{4pt}
\titlespacing*{\subsection}{0pt}{15pt}{1pt}

\fancyfoot{}
\fancyfoot[LO,RE]{\vspace{-7.1pt}\includegraphics[height=9pt]{head_foot/LF}}
\fancyfoot[CO]{\vspace{-7.1pt}\hspace{13.2cm}\includegraphics{head_foot/RF}}
\fancyfoot[CE]{\vspace{-7.2pt}\hspace{-14.2cm}\includegraphics{head_foot/RF}}
\fancyfoot[RO]{\footnotesize{\sffamily{1--\pageref{LastPage} ~\textbar  \hspace{2pt}\thepage}}}
\fancyfoot[LE]{\footnotesize{\sffamily{\thepage~\textbar\hspace{3.45cm} 1--\pageref{LastPage}}}}
\fancyhead{}
\renewcommand{\headrulewidth}{0pt} 
\renewcommand{\footrulewidth}{0pt}
\setlength{\arrayrulewidth}{1pt}
\setlength{\columnsep}{6.5mm}
\setlength\bibsep{1pt}

\makeatletter 
\newlength{\figrulesep} 
\setlength{\figrulesep}{0.5\textfloatsep} 

\newcommand{\topfigrule}{\vspace*{-1pt}%
\noindent{\color{cream}\rule[-\figrulesep]{\columnwidth}{1.5pt}} }

\newcommand{\botfigrule}{\vspace*{-2pt}%
\noindent{\color{cream}\rule[\figrulesep]{\columnwidth}{1.5pt}} }

\newcommand{\dblfigrule}{\vspace*{-1pt}%
\noindent{\color{cream}\rule[-\figrulesep]{\textwidth}{1.5pt}} }

\makeatother

\newcommand*\mycommand[1]{\texttt{\emph{#1}}}
\newcommand{\yamazaki}[1]{\textcolor{cyan}{#1}}

\twocolumn[
  \begin{@twocolumnfalse}
{\includegraphics[height=30pt]{head_foot/journal_name}\hfill\raisebox{0pt}[0pt][0pt]{\includegraphics[height=55pt]{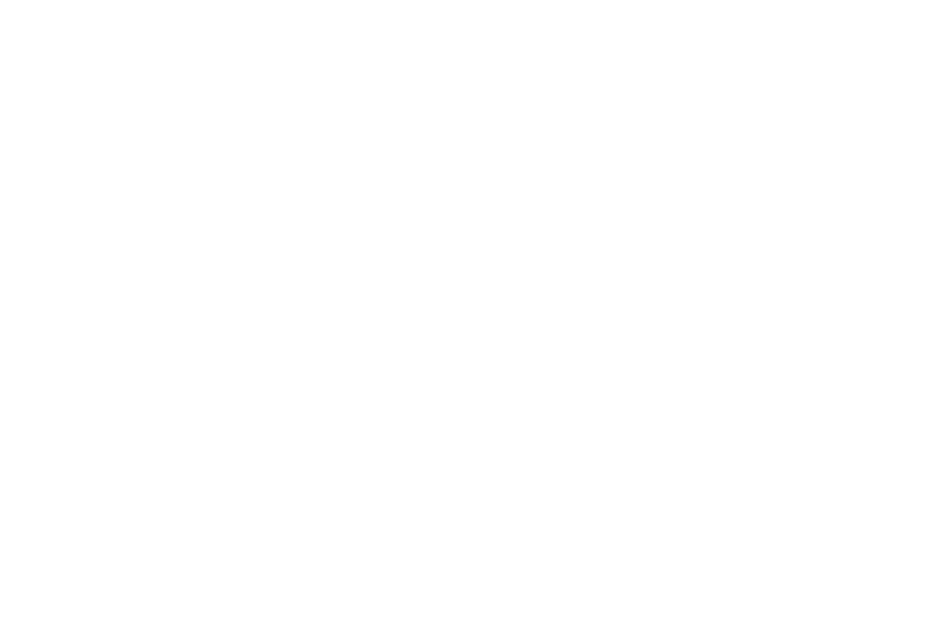}}\\[1ex]
\includegraphics[width=18.5cm]{head_foot/header_bar}}\par
\vspace{1em}
\sffamily
\begin{tabular}{m{4.5cm} p{13.5cm} }

\includegraphics{head_foot/DOI} & \noindent\LARGE{\textbf{Electronic Supplementary Information: Population Trap in X-ray Induced Ultrafast Nonadiabatic Dynamics of Tropone Probed at the O(1\textit{s}) pre-edge}} \\
\vspace{0.3cm} & \vspace{0.3cm} \\

 & \noindent\large{Kaoru Yamazaki,$^{\ast}$\textit{$^{a}$} and Katsumi Midorikawa\textit{$^{a}$}} \\

\includegraphics{head_foot/dates} & \noindent\normalsize{} \\

\end{tabular}

 \end{@twocolumnfalse} \vspace{0.6cm}
]


\renewcommand*\rmdefault{bch}\normalfont\upshape
\rmfamily
\section*{}
\vspace{-1cm}


\footnotetext{\textit{$^{a}$~Attosecond Science Research Team, Extreme Photonics Research Group, RIKEN Center for Advanced Photonics, RIKEN, 2-1 Hirosawa, Wako, Saitama, 351-0198, Japan. Fax: +81 48-462-4682; Tel: +81-50-3502-5574; E-mail: kaoru.yamazaki@riken.jp}}




\section{Details of computational methods}

\subsection{Auger decay probability from two-population analysis}
Given that an core ionization creates an core-hole state $\ket{\Psi_i^{1s}}$ whose core-hole is located at the $i$-th $1s$ molecular orbital (MO) located on the atoms A and it rapidly undergoes Auger decay. In the single-electron configuration limit, the $\ket{\Psi_i^{1s}}$ can be calculated by a $\Delta$SCF procedure and the relative Auger amplitude $t_{ab}$ can be written as a function of the L\"owdin electron populations of MOs $a$ and $b$ on the core-hole atoms A, $Q_a(\mathrm{A})$ and $Q_a(\mathrm{A})$  with assuming that the contribution from the continuum photoelectron and Auger electron is constant for all decay channels.\cite{mitani2003} The representations of $t_{ab}$ for singlet configuration are\cite{mitani2003}

%
\begin{align}
    \label{eq_auger_singlet_ab}
    t_{ab} (\mathrm{A}) & \approx \sqrt{\frac{1}{2}} (Q_a (\mathrm{A}) + Q_b (\mathrm{A})), & (a \neq b ) \\
    \label{eq_auger_singlet_aa}
    t_{aa} &\approx Q_a(\mathrm{A}), &(a = b)
\end{align}
and that for triplet one is\cite{mitani2003}
\begin{align}
    \label{eq_auger_triplet_ab}
    t_{ab} (\mathrm{A}) & \approx \sqrt{\frac{3}{2}} (Q_a (\mathrm{A}) - Q_b (\mathrm{A})). & (a \neq b )
\end{align}

In general, the electronic wave function of the dicationic state $\ket{\Psi^{q=2}}$ can be represented as the linear combination of the all possible two hole-electron configurations.\cite{mitani2003} The wave function within the two-hole configuration interaction theory on the top of $\ket{\Psi_i^{1s}}$ is written using the second quantization notation as
\begin{equation}
\label{eq_2hci}
    \ket{\Psi^{q=2}_i} = \sum_{a \le b} c_a c_b c_{1s}^\dagger\ket{\Psi_i^{1s}},
\end{equation}
where $c_a$, $c_b$, $c_{1s}^\dagger$ are the annihilation operater for an electron in the MO $a$ and $b$, and the creation operator for that in  the core-hole $1s$ MO. We further included the MO relaxation effect by taking the overlap between the $\ket{\Psi^{q=2}_i}$ and the $\nu$-th dication wave function calculated by the state-averaged complete active space self consistent active field methods averaged over all two-hole electronic state (SA-2h-CASSCF) for given spin-multiplicity $\ket{\Psi^{q=2}_\nu}$. The final expression for the relative Auger transition intensity from  $\ket{\Psi_i}$ to $\ket{\Psi^{q=2}_\nu}$ is written as\cite{inhester2018}
\begin{equation}
    \label{eq_auger_final}
    I_{i \rightarrow \nu} 
    = 2 \pi \left| \sum_{a \le b} \bra{\Psi^{q=2}_\nu} c_a c_b c_{1s}^\dagger\ket{\Psi_i^{1s}}t_{ab} \right|^2,
\end{equation}
The normalized cumulative relative intensity (Auger spectrum) for the $\ket{\Psi^{q=2}_\nu}$ as a function of its two-hole electron binding energy $E_\nu$ represents the initial population for the subsequent nonadiabatic reaction dynamics simulations:
\begin{align}
    \label{eq_auger_normalized}
    I_\nu(E_\nu) &= \frac{\sum_{i}I_{i \rightarrow \nu} }{\sum_{i,\nu} I_{i \rightarrow \nu} },\\
    \label{eq_initial_population}
    \tilde{P}_\nu(t=0) &= I_\nu.
\end{align}

\subsection{Nonadiabatic reaction dynamics}

We performed surface hopping molecular dynamics simulations on the NRD dynamics of tropone dications using an Tully's fewest switching algorithm combined with a linear vibronic coupling (LVC) model potential energy surfaces (PESs) \cite{Plasser2018,heim2020,zobel2021} (LVC-MD). 
We included 210 singlet and 190 triplet two-hole states, and the all 36 vibrational degrees of freedom. The details of the evaluations of the LVC parameters are described in section 1.3.
Trajectories were set up using 1000 initial coordinates and momenta from the Wigner
distribution of the ground-state potential energy surface at the temperature of 108 \(^\circ\)C according to an gas phase electron diffraction experiment by Ogasawara \textit{et al.} \cite{Ogasawara1972}. The initial dication states for 200 trajectories were selected according to the relative normal Auger probability\cite{li2015_prl} initiated by the core ionization at the Carbon $1s$ edge (See Figure 1 in the main text). 

%
%


The trajectories were then propagated on the LVC potential energy surfaces until they encountered to a surface hop at a dissociative crossing point of the LVC potential energy surfaces or trajectory lengths reached 1000 fs.  Each trajectory was stopped if it reaches the dissociative states where any bond-length elongates to the threshold bond-length $R_\text{th}=$ 3.4 \AA, which corresponds to the double of the Van der Waals radius of carbon atom, and thereafter regarded as \textit{vibrationally hot} molecule.

The settings for the LVC-MD are as follows:
We used a nuclear time step of 0.1 fs and an electronic time step of 0.004 fs An energy-based decoherence correction with a constant of 0.1 Hartree \cite{granucci2010} was used. The kinetic energy was adjusted by rescaling the velocity vectors. The time derivative couplings to evaluate the surface hopping probabilities were approximated using the wave function overlaps.
A local diabatization method\cite{May2014} was applied to evaluated the $\{ \tilde{P}_\nu(t)\}$ for the LVC-MD.

\subsubsection{Linear vibronic coupling (LVC) model}

In an LVC model\cite{Plasser2018}, the PESs are approximated via the ground-state PES $V_0$ and the first-order (linear) vibronic coupling term $\mathbf{W}$ on the basis of the mass-frequency scaled normal mode coordinate $\mathbf{x} =[x_1,x_2,\dots,x_l,\dots, x_{3N'-6}]$,
\begin{equation}
V(\mathbf{x}) = V_0(\mathbf{x})\mathbf{1} + \mathbf{W}(\mathbf{x}).    
\end{equation}

The ground-state PESs for $N'$-atom molecules are approximated as harmonic oscillators with frequencies $\omega_l$ as
\begin{equation}
   V_0(\mathbf{x}) = \sum_{l = 1}^{3N'-6} 
   \frac{\hbar \omega_l}{2}x_l^2.
\end{equation}

The coupling matrix $\mathbf{W}(\mathbf{x})$ serves to create the individual potential energy surfaces from the
reference potential, and consists of constant and linear terms
\begin{equation}
    W_{nn}(\mathbf{x}) = E_\nu^{(n)} + \sum_{l = 1}^{3N'-6} \kappa_l^{(n)} x_l,
\end{equation}
and 
\begin{equation}
    W_{mn}(\mathbf{x}) = \sum_{l = 1}^{3N'-6} L_l^{(n,m)} x_l
\end{equation}
where $E_\nu^{(n)}$ are the two-hole binding energies at $\mathbf{x}_0=\mathbf{R}_0$, while $\kappa_l^{(n)} x_l$ and $L_l^{(n,m)}$ are the intrastate and interstate coupling elements for the normal mode coordinate $x_l$. The coupling elements were obtained numerically from SA-2h-CASSCF calculations (see below) on geometries displaced by ±0.05 units from $\mathbf{x}_0$ for each normal mode. The intrastate coupling $\kappa_l^{(n)} x_l$ values were obtained as numerical gradients and the interstate couplings $L_l^{(n,m)}$ were obtained from the change in the wavefunction overlaps. The spin–orbit couplings (SOCs) in the LVC model potential were approximated by constant SOCs obtained at $\mathbf{x}_0$.

\subsection{Quantum chemistry methods}

In this study, we consequently used the optimized geometry in the the neutral grand state of tropone at the density fitted localized  M\"oller–Plesset second order perturbation (DF-LMP2) theory\cite{Werner2020} combined with the Def2-TZVP basis set\cite{Weigend2006}. The convergence of geometry optimization was confirmed by the normal mode analysis at the same level of theory.  

For the Auger spectrum calculation and LVC-parameter evaluation, we used the SA-2h-CASSCF method including all two-hole states represented by 20 valence active MO (the [2h, 20o] RAS2 space) combined with ANO-RCC-VTZP basis set. 210 singlet and 190 triplet states were included, respectively. The SOCs were evaluated by the Breit-Pauli SOC operator with the mean-field approximation combined with the restricted active space state interaction (RAS-SI) method\cite{Hess1996,Malmqvist2002}. 

The $\epsilon_{\nu \mu} (\mathbf{R}_0)$ were calculated by the SA-2h-CASSCF/ANO-RCC-VTZP method with extending the RAS2 space to 20 valence MO and O($1s$) MO, and by RAS-SI method. The photoabsorption cross sections for the C$($1s$) \rightarrow V^{-1}$ transitions of the doublet monocations at the Franck-Condon geometry were also calculated at the SA-1h-CASSCF/ANO-RCC-VTZP method which includes 20 valence MO and 7 C($1s$) MO to the RAS2 space combined with the RAS-SI method. We included all 27 monocationic states to this calculation.
We did not used point group symmetry for the SA-1h/2h-CASSCF calculations above.

For the x-ray photoelectron spectra (XPS) of the lowest energy low spin state of Tr$^{q+} $ $(q = 0,1,2)$ at the C($1s$) and O($1s$) edges including low-lying shape-up states. We used non-relativistic multi-state restricted active space second order perturbation theory (MS-RASPT2) \cite{Malmqvist2008} with use of $C_{2v}$ point group symmetry. We used the Supporo-2012-TZP basis set for C and O atoms, and the Supporo-2012-DZP basis set for H atoms\cite{Noro2012}. We created reference wave functions for initial and final states by single-state and state-average restricted active space self-consistent field (RASSCF) method, respectively. We included four $\pi$, four $\pi^*$, and one $n$ orbitals, and the all $10-q$ electrons in the RAS2 space ([$(10-q)\text{e}, 9\text{o}$] RAS2 space) as shown in Figure S1. We added all C($1s$) or O($1s$) orbitals to RAS1 space for C($1s$) and O($1s$) XPS calculation, respectively. We created one core-hole by the HEXS keyword and super-symmetry for final core-hole state calculations\cite{aquilante2020}. 
The number of core-hole states included in the final state calculations were summarized in Table S1. The SA-RASSCF calculations and subsequent MS-RASPT2 calculations were performed for each irreducible representation group. We used the IPEA shift of 0.25 Hartree and imaginary shift of 0.20 Hartree to reduce intruder problems on MS-RASPT2 calculations. The relative intensity of the XPS  was evaluated by using the norm of Dyson orbitals \textit{via} the RAS-SI method.

We used Molpro 2019.2 quantum chemistry package for the geometry optimization and normal mode analysis\cite{Werner2020}. The OpenMolcas 18.09 and 21.10 quantum chemistry packages\cite{Galvan2019_jctc,aquilante2020} were used for all SA-2h/1h-CASSCF and MS-RASPT2 calculations, respectively. Nonadiabatic reaction dynamics simulations were performed with the SHARC   2.1 program\cite{Mai2018}.

\begin{figure}[h]
  \includegraphics[width=1.0\linewidth]{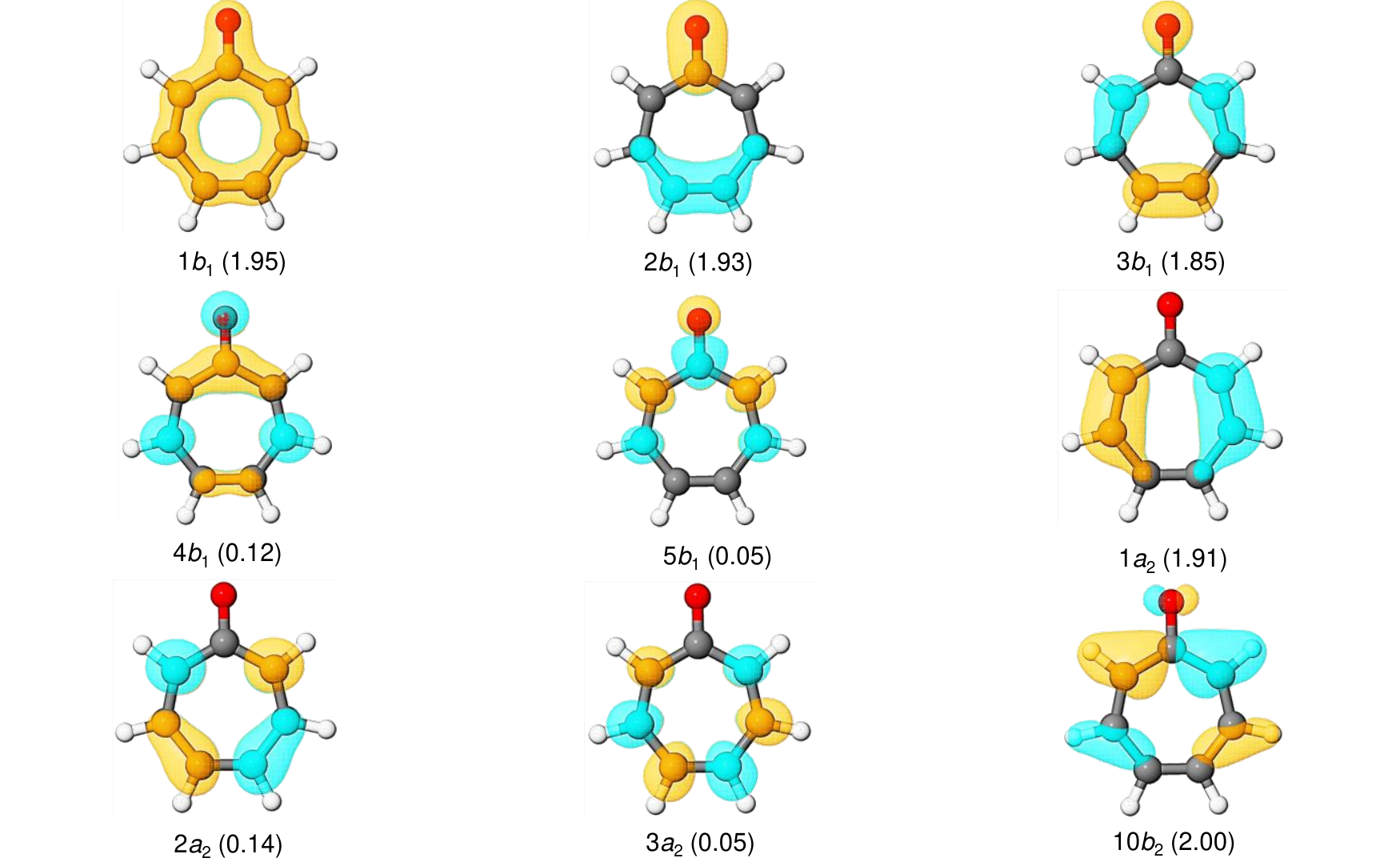}
  \caption*{Fig. S1 The natural orbitals included in the RAS2 space for the XPS calculations for Tr at the MS-RASPT2/Sapporo-2012 (C, O: TZP, H: DZP) level of theory. The occupation numbers of the natural orbitals are indicated in the parenthesis.}
\label{fig:ras2}
\end{figure}

\begin{table}[h]
  \caption*{Table S1 The number of core-hole states in each irreducible representation for the XPS calculations on Tr$^{q+} (q = 0,1,2)$ at the C($1s$) and O($1s$) edges by means of the MS-RASPT2 level of theory.}
  \label{table:no_final_states_xps}
  \centering
  \begin{tabular}{lrlrrrrr}
    \hline
    Absorption & & Spin  &  &  &  &  &  \\
    edge &$q$& multiplicity & $A_1$ & $B_1$ & $A_2$ & $B_2$ & Total \\
    \hline
    C$($1s$)$ & 0 & Doublet & 60& 14& 14& 59& 147\\
    C$($1s$)$ & 1 & Singlet & 22& 68& 65& 20& 175\\
    C$($1s$)$ & 1 & Triplet & 22& 68& 65& 20& 175\\
    C$($1s$)$ & 2 & Doublet & 57& 17& 18& 57& 149\\
    O$($1s$)$ & 0 & Doublet & 9& 2& 2& 8& 21\\
    O$($1s$)$ & 1 & Singlet & 2& 11& 8& 4& 25\\
    O$($1s$)$ & 1 & Triplet & 2& 11& 8& 4& 25 \\
    O$($1s$)$ & 2 & Doublet & 9& 2& 3& 7& 21\\
    \hline
  \end{tabular}
 \end{table}
\section{Ionization kinetics including sequential double ionization process of Tr}

\subsection{Photo-ions yields for the pump pulse irradiation in saturable regime}

We first consider the TR-XAS signal including one- and two-photon sequential ionization processes by the pump pulse and one-photon ones by the probe pulse.

The pump pulse can trigger the core ionization and subsequent normal Auger decay for Tr at the C$(1s)$ edge ($C$), valence ionization ($V^{-1}$) for Tr$^{q+} (q = 0,1,2)$, and core excitation from C$(1s)$  edge to valence holes and subsequent resonance Auger decay ($ \text{C}(1s) \rightarrow V$) for Tr$^{+}$ and Tr$^{q+}$. The possible two ionization pathways by the pump pulse are 
\begin{equation}
    \ce{Tr}  \xrightarrow{C} \ce{Tr}^{2+} \xrightarrow{V^{-1} \text{ or } \text{C}(1s) \rightarrow V} \ce{Tr}^{3+}, \label{eq:pump_signal}
\end{equation}
and
\begin{equation}
    \ce{Tr}  \xrightarrow{V^{-1}} \ce{Tr}^{+} \xrightarrow{V^{-1} \text{ or } \text{C}(1s) \rightarrow V} \ce{Tr}^{2'+}. \label{eq:pump_bg}
\end{equation}
The $C$ process from Tr to Tr$^{2+}$ contributes to the signal and the others do to background noise. As the photo-ionization and absorption cross section are almost independent from $q$ \cite{guichard2013}, the yield of Tr$^{q+}$ ($q=0,1,2,2'$) after pump irradiation can be written as functions of $I_\text{pu}$ and $\hbar \omega_\text{pu}$ as
 \begin{align}
    \label{eq_q0_pu}
    [\ce{Tr}](I_\text{pu}, \hbar \omega_\text{pu}) 
    &= \exp \left( -\lambda_{V}(I_\text{pu}, \hbar \omega_\text{pu})
    -\lambda_{C}(I_\text{pu}, \hbar \omega_\text{pu})\right),\\
    \label{eq_q1_pu}
    [\ce{Tr}^{+}] (I_\text{pu}, \hbar \omega_\text{pu}) 
    &= \frac{\sigma_{V}(\hbar \omega_\text{pu})}{\sigma_C(\hbar \omega_\text{pu}) - \sigma_{ \text{C}(1s) \rightarrow V}(\hbar \omega_\text{pu})} \nonumber \\
    &\times [
        \exp \left( -\lambda_V(I_\text{pu}, \hbar \omega_\text{pu}) - \lambda_{ \text{C}(1s) \rightarrow V}(I_\text{pu}, \hbar \omega_\text{pu}) \right) \nonumber\\
        &~-\exp\left( -\lambda_V(I_\text{pu}, \hbar \omega_\text{pu}) - \lambda_C(I_\text{pu}, \hbar \omega_\text{pu}) \right)],\\
    \label{eq_q2_pu}
    [\ce{Tr}^{2+}] (I_\text{pu}, \hbar \omega_\text{pu}) 
    &= 
        \exp \left( -\lambda_V(I_\text{pu}, \hbar \omega_\text{pu})  \right) \nonumber\\
        &~-\exp\left( -\lambda_V(I_\text{pu}, \hbar \omega_\text{pu}) - \lambda_C(I_\text{pu}, \hbar \omega_\text{pu}) \right),\\
    \label{eq_q2p_pu}
    [\ce{Tr}^{2'+}] (I_\text{pu}, \hbar \omega_\text{pu}) 
    &= \frac{\sigma_{V}(\hbar \omega_\text{pu})}{\sigma_V(\hbar \omega_\text{pu}) + \sigma_C(\hbar \omega_\text{pu})}  \nonumber\\
    &\times 
    \bigg[ 1 + 
     \frac{1}{\sigma_C(\hbar \omega_\text{pu}) - \sigma_{ \text{C}(1s) \rightarrow V}(\hbar \omega_\text{pu})} \nonumber\\
     &\times \{(\sigma_V(\hbar \omega_\text{pu}) + \sigma_{ \text{C}(1s) \rightarrow V}(\hbar \omega_\text{pu})) \nonumber\\
     & \times \exp \left( -\lambda_V(I_\text{pu}, \hbar \omega_\text{pu}-\lambda_C(I_\text{pu}, \hbar \omega_\text{pu}) \right) \nonumber\\
        &-(\sigma_V(\hbar \omega_\text{pu}) + \sigma_C(\hbar \omega_\text{pu})) \nonumber\\
         & \times \exp \left( -\lambda_V(I_\text{pu}, \hbar \omega_\text{pu}-\lambda_{ \text{C}(1s) \rightarrow V}(I_\text{pu}, \hbar \omega_\text{pu}) \right)
        \} 
    \bigg],\\
    \label{eq_q3p_pu}
    [\ce{Tr}^{3+}] (I_\text{pu}, \hbar \omega_\text{pu}) 
    &= \frac{\sigma_{C}(\hbar \omega_\text{pu})}{\sigma_V(\hbar \omega_\text{pu}) + \sigma_C(\hbar \omega_\text{pu})}  \nonumber\\
    &\times 
    \bigg[ 1 + 
     \frac{1}{\sigma_C(\hbar \omega_\text{pu})} \nonumber\\
     &\times \{\sigma_V(\hbar \omega_\text{pu}) \nonumber\\
     & \times\exp \left( -\lambda_V(I_\text{pu}, \hbar \omega_\text{pu})-\lambda_C(I_\text{pu}, \hbar \omega_\text{pu}) \right) \nonumber\\
        &-(\sigma_V(\hbar \omega_\text{pu}) + \sigma_C(\hbar \omega_\text{pu})) \nonumber\\
         &\times \exp \left( -\lambda_V(I_\text{pu}, \hbar \omega_\text{pu} \right)
        \} 
    \bigg],
\end{align}
where $\sigma_V(\hbar \omega_\text{pu})$, $ \sigma_C(\hbar \omega_\text{pu})$, and $ \sigma_{ \text{C}(1s) \rightarrow V}(\hbar \omega_\text{pu})$ represent the photoionization cross sections of the $V^{-1}$, $C$, and $ \text{C}(1s) \rightarrow V$ ionization processes as functions of $\hbar \omega_\text{pu}$, respectively. $\lambda_{V}(I_\text{pu},\hbar \omega_\text{pu})$ and $\lambda_{C}(I_\text{pu}, \hbar \omega_\text{pu})$ mean the number of absorbed photons during the pump irradiation for the ionization processes $V^{-1}$ and $C$, respectively.  For the Gaussian pump pulse with a peak intensity of $I_\text{pu}$ and full width at half maximum (FWHM)of $s_\text{pu}$, $\lambda_{A_\text{pu}}(I_\text{pu}, \hbar \omega_\text{pu})~ (A_\text{pu} = V^{-1}, C, X\rightarrow V)$ is written as
\begin{equation}
    \label{eq:def_lambda_pu}
    \lambda_{A_\text{pu}}(I_\text{pu}, \hbar \omega_\text{pu}) 
    = \sigma_{A_\text{pu}}(\hbar \omega_\text{pu}) \frac{I_\text{pu}}{\hbar \omega_\text{pu}}
    \sqrt{
    \frac{\pi}{4\ln2}
    } s_\text{pu}.
\end{equation}
%

\subsection{Yield of Auger electron generated by the probe pulse in linear regime}

One photon absorption from the probe pulse at the $Y$-pre-edge of Tr$^{2+}$ excites core electron at $Y$-edge to the valence hole. The core-excited Tr$^{2+}$ which is rapidly ionized into Tr$^{3+}$ via resonance Auger relaxation emits one fast Auger electron e$^{-}_\text{Auger}$,
\begin{equation}
    \ce{Tr}^{2+}  \xrightarrow{ \text{O}(1s) \rightarrow V} \ce{Tr}^{3+} + \ce{e}^{-}_\text{Auger}. \label{eq:probe_bg}
\end{equation}

The TR-XAS measurement using the Auger electron yield method \cite{Palacios2020,Driver2020,Barillot2021,Li2021_fd,Li2022_science,Khalili2021,Schwickert2022_sciadv} selectively counts the number of e$^{-}_\text{Auger}$ as a function of $\hbar \omega_\text{pr}$ and $\Delta t$ by using electron spectrometers. This scheme can eliminate the photoelectrons from core ionization and $V^{-1}$ processes, and the Auger electrons from the  $C$ process of Tr$^{q+} (q = 0,1,2,2',3)$ by the pump and probe pulses. The background sources can come from the resonance Auger electrons from the $ \text{O}(1s) \rightarrow V$  excitation process of Tr$^+$ and Tr$^2{+}$, and normal Auger electron from O$(1s)^{-1}$ core ionization of Tr. The contribution of Tr$^{2+'}$ is negligibly small since $\sigma_{V}(\hbar \omega_\text{pu})$ is smaller than $\sigma_{C}(\hbar \omega_\text{pu})$ by a magnitude nearby the C$(1s)$ edges of Tr and Tr$^+$.
 
When the temporal overlap between the pump and probe pulse is negligibly small, the Auger electron yield $N_\text{e}{(q)} $ originated from Tr$^{q+} (q = 0,1,2,2')$ can be written as,
\begin{align}
    \label{eq:electron_yield}
    &N_\text{e}(q, I_\text{pu}, \hbar \omega_\text{pu},I_\text{pr}, \hbar \omega_\text{pr}) \notag\\
    &= \theta (q, \hbar \omega_\text{pr})[\ce{Tr}^{q+}](I_\text{pu}, \hbar \omega_\text{pu})\frac{\sigma_{A_\text{pr}}(q, \hbar \omega_\text{pr})}{\sigma_\text{tot}(q,\hbar \omega_\text{pr})} \notag\\
    & \times \left[
        1-\exp(-\lambda_\text{tot}(q, I_\text{pr}, \hbar \omega_\text{pr}))
    \right].
\end{align}
Here, $\theta (q, \hbar \omega_\text{pr})$ describes the fraction of bright states for Tr$^{q+}$ at $\hbar \omega_\text{pr}$ ($\theta (q= 0, \hbar \omega_\text{pr}) = 1$ and $0\le \theta (0 < q, \hbar \omega_\text{pr}) \le 1$).  $\sigma_{A_\text{pr}}(q, \hbar \omega_\text{pr})$ represents the mean photoabsorption crossection for the O$(1s) \rightarrow V$ excitation and subsequent resonance Auger decay ($q=1, 2,2'$), and the O$(1s)^{-1}$ core ionization and subsequent oxygen $KLL$ Auger decay ($O$) for $q=0$, respectively.
$\sigma_\text{tot}(q,\hbar \omega_\text{pr}) \equiv \sigma_{A_\text{pr}}(q, \hbar \omega_\text{pr}) + \sigma_{V}(\hbar \omega_\text{pr})+ \sigma_{C}(\hbar \omega_\text{pr})$  is the total photoionization cross section for Tr$^{q+}$ as a function of $\hbar \omega_\text{pr}$, and $\lambda_\text{tot}(q, I_\text{pr}, \hbar \omega_\text{pr})$ is the total number of absorbed photon in the probe irradiation. 
For the Gaussian probe pulse with a peak intensity of $I_\text{pr}$  and FWHM of $s_\text{pr}$, $\lambda_\text{tot}(q, I_\text{pr}, \hbar \omega_\text{pr})$ is written as
\begin{equation}
    \label{eq:def_lambda_pr}
    \lambda_\text{tot}(q, I_\text{pr}, \hbar \omega_\text{pr}) 
    = \sigma_\text{tot}(q, \hbar \omega_\text{pr}) \frac{I_\text{pr}}{\hbar \omega_\text{pr}}
    \sqrt{
    \frac{\pi}{4\ln2}
    } s_\text{pr}.
\end{equation}
In the liner probe regime, eq \eqref{eq:electron_yield} is finally simplified to
\begin{align}
        \label{eq:electron_yield_liner}
    &N_\text{e}(q, I_\text{pu}, \hbar \omega_\text{pu},I_\text{pr}, \hbar \omega_\text{pr}) 
    \approx 
    \theta (q, \hbar \omega_\text{pr})
    [\ce{Tr}^{q+}](I_\text{pu}, \hbar \omega_\text{pu}) \notag\\
    & \times \sigma_{A_\text{pr}}(q, \hbar \omega_\text{pr})
    \frac{I_\text{pr}}{\hbar \omega_\text{pr}}
    \sqrt{
    \frac{\pi}{4\ln2}
    } s_\text{pr}.
\end{align}

The S/B $\eta$ can be defined as the ratio of $ N_\text{e}(q=2, I_\text{pu}, \hbar \omega_\text{pu},I_\text{pr}, \hbar \omega_\text{pr})$ and the sum of the electron yields originated from other O$(1s)$-edge (resonance) Auger processes,
\begin{equation}
\label{eq:S/B_def_text}
    \eta( I_\text{pu}, \hbar \omega_\text{pu},I_\text{pr}, \hbar \omega_\text{pr}) 
    \equiv \frac{
    N_\text{e}(q=2, I_\text{pu}, \hbar \omega_\text{pu},I_\text{pr}, \hbar \omega_\text{pr})}{
    \sum_{q=0,1,2'}N_\text{e}(q, I_\text{pu}, \hbar \omega_\text{pu},I_\text{pr}, \hbar \omega_\text{pr})}.
\end{equation}

\subsection{Photoionization cross sections}
Core and valance photoionization cross section  $\left\{\sigma_{A} \middle| A = \ce{C}(1s), \ce{O}(1s), V \right \}$ of Tr$^{q+}$ ($q=0,1,2$) were calculated from the atomistic calculations by Yeh. et al.\cite{Yeh1985} We used the values of $\hbar \omega = 300$ eV for the representative values of $\{\sigma_{A}\}$ above the C($1s$) edge  and $\hbar \omega = 542 $ eV for O($1s$) (pre-) edge. The photoabosorption cross sections  of C$($1s$)\rightarrow V^{-1}$ of Tr$^+$ and O$($1s$)\rightarrow V^{-1}$ of Tr$^{2+}$ were calculated by averaging over the photoabsorption cross sections of the corresponding optically allowed core-to-valence transitions whose oscillator strengths are larger than 0.01\cite{Couto2020,Zinchenko2021}. We assumed that the all core-to-valence excited states rapidly undergoes the resonance Auger process and Tr$^{q+}$ is ionized to Tr$^{(q+1)+}$. We also given that these photoionization cross sections do not depend on $q$\cite{guichard2013}.

\begin{table}
  \caption*{Table S2 The photo-ionization cross sections $\sigma_{A_\text{pu}}(\hbar \omega_\text{pu})$ and $\sigma_{A_\text{pr}}(\hbar \omega_\text{pr})$ used in the calculation in $10^{-18}$ cm$^2$. }
  \label{table:tau_trxas_vs_population}
  \centering
  \begin{tabular}{lrr}
    \hline
    $A$  & $\sigma_{A_\text{pu}}(\hbar \omega_\text{pu})$ & $\sigma_{A_\text{pr}}(\hbar \omega_\text{pr})$ \\
    \hline
    $V$            & 0.45$^a$         & 0.18$^a$\\
    C$($1s$)$        & 6.01$^a$         & 1.55$^a$\\
    C$($1s$) \rightarrow V$       & 1.69$^b$         & ---\\
    O$($1s$)$        & ---         & 0.51$^a$\\
    O$($1s$) \rightarrow V$       & ---         & 2.35$^c$\\
    \hline
  \end{tabular}
 \\ $^a$Yeh et al.\cite{Yeh1985}; \\$^b$ SA-1h-CASSCF/ANO-RCC-VTZP//DF-LMP2/Def2-TZVP;\\ $^c$SA-2h-CASSCF/ANO-RCC-VTZP//DF-LMP2/Def2-TZVP.
 \end{table}

 \section{X-ray photoelectron spectra of Tr$^{q+} (q = 0,1,2)$ at the O($1s$) edge}
 
 \begin{figure}[h]
 \centering
  \includegraphics[width=1.0\linewidth]{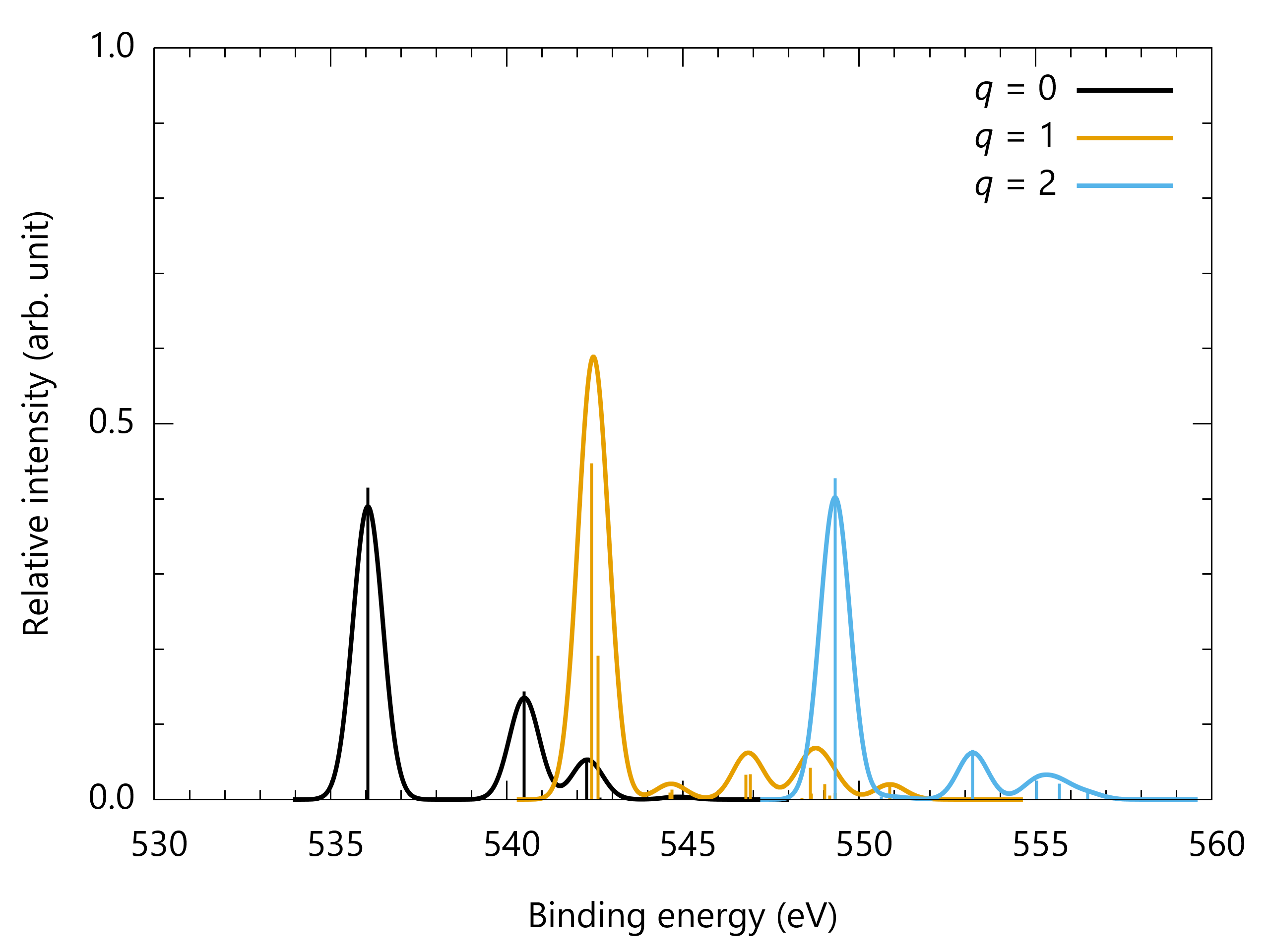}
  \caption*{Fig. S2 The calculated  XPS spectra of Tr$^{q+} (q = 0,1,2)$ at the O($1s$) edge. Results at the MS-RASPT2/Sapporo-2012 (C,O: TZP, H: DZP)//DF-LMP2/Def2-TZVP level of theory. The calculated spectra were broaden by a Gaussian function of FWHM  = 1.0 eV.}
\label{fig:o1s_xps}
\end{figure}


\bibliography{library} 
\bibliographystyle{rsc} 
\balance